\documentclass[graybox]{svmult}

\usepackage{mathptmx}  
\usepackage{helvet}         
\usepackage{courier}        
\usepackage{type1cm}      
\usepackage{makeidx}         
\usepackage{graphicx}        
\usepackage{multicol}        
\usepackage[bottom]{footmisc}

\usepackage[english]{babel}
\usepackage{graphicx}
\usepackage{amssymb,amsmath}
\usepackage{bm}
\usepackage{fancyhdr}
\usepackage[latin1]{inputenc}
\usepackage{aas_macros}
\DeclareGraphicsExtensions{.jpg,.pdf,.eps}

\makeindex             

\def\msun{{M}_\odot}
\def\thesection{\arabic{section}}
\def\beq{\begin{equation}}
\def\eeq{\end{equation}}
\def\beqn{\begin{eqnarray}}
\def\eeqn{\end{eqnarray}}

\def\gcm{g~cm$^{-3}$}

\def\P3{{\cal P}_t}
\def\J3{{\cal J}}
\def\T3{{\cal T}}

\def\bra{\langle}
\def\ket{\rangle}

\def\beq{\begin{equation}}
\def\eeq{\end{equation}}
\newcommand{\be}{\begin{eqnarray}}
\newcommand{\ee}{\end{eqnarray}}
\def\bar{\begin{array}[b]}
\def\barc{\begin{array}}
\def\bart{\begin{array}[t]}
\def\ear{\end{array}}
\def\le#1{\label{eq:#1}}

\begin{document}

\title*{Nuclear Equation of state for Compact Stars and Supernovae}
\author{G. Fiorella Burgio and Anthea F. Fantina}
\institute{G. Fiorella Burgio \at Istituto Nazionale di Fisica Nucleare, Sez. di Catania, Via S. Sofia 64, 95123 Catania, Italy, \email{fiorella.burgio@ct.infn.it}
\and Anthea F. Fantina \at Grand Acc\'el\'erateur National d'Ions Lourds (GANIL), CEA/DRF - CNRS/IN2P3, Bvd Henri Becquerel, 14076 Caen, France, \email{anthea.fantina@ganil.fr}}
%
%
\maketitle

\abstract{
The equation of state (EoS) of hot and dense matter is a fundamental input to describe static and dynamical properties of neutron stars, core-collapse supernovae and binary compact-star mergers. 
We review the current status of the EoS for compact objects, that have been studied with both ab-initio many-body approaches and phenomenological models. 
We limit ourselves to the description of EoSs with purely nucleonic degrees of freedom, disregarding the appearance of strange baryonic matter and/or quark matter. 
We compare the theoretical predictions with different data coming from both nuclear physics experiments and astrophysical observations.
Combining the complementary information thus obtained greatly enriches our insight into the dense nuclear matter properties.
Current challenges in the description of the EoS are also discussed, mainly focusing on the model dependence of the constraints extracted from either experimental or observational data, the lack of a consistent and rigorous many-body treatment at zero and finite temperature of the matter encountered in compact stars (e.g. problem of cluster formation and extension of the EoS to very high temperatures), the role of nucleonic three-body forces, and the dependence of the direct URCA processes on the EoS.
}

\section{Introduction}
\label{sec:intro}


An equation of state (EoS) is a relation between thermodynamic variables describing the state of matter under given physical conditions.
Independent variables are usually the particle numbers $N_i$, the temperature $T$, and the volume $V$; alternatively, one can use the particle number densities $n_i=N_i/V$, the corresponding particle number fractions being $Y_i=n_i/n_B=N_i/N_B$ (e.g. $Y_e=N_e/N_B$ for the electron fraction, $N_B$ and $N_e$ being the baryon and electron number, respectively).
In addition, conservation laws hold, so that there are conserved quantities such as the total baryon number, the total electric charge number, and the total lepton number.

In astrophysics, EoSs are usually implemented in hydrodynamic (or hydrostatic) models that describe the evolution (or the static structure) of the macroscopic system.
An EoS can be determined if the system is in thermodynamic equilibrium, i.e. if thermal, mechanical, and chemical equilibrium are achieved.
In particular, the latter generally is not attained in main sequence stars or in explosive nucleosynthesis.
In these scenarios, a full reaction network that takes into account the reaction cross sections of the species present in the medium has to be considered.
Otherwise, if the timescale of the nuclear reactions is much shorter than the dynamic evolution timescales, nuclear statistical equilibrium (NSE) can be assumed.
This situation is typically achieved for temperatures $T \gtrsim 0.5$~MeV \cite{iliadis2007}.
On the other hand, weak interactions are not generally in equilibrium.
Specifically, in core-collapse supernovae (CCSNe), the electron-capture reaction, $p + e^- \rightarrow n + \nu_e$, is not in equilibrium for baryon densitites below $n_B \approx 10^{-3} - 10^{-4}$~fm$^{-3}$ (or equivalently, for mass-energy densities below $\rho_B \approx $~few~$10^{11}$~g~cm$^{-3}$).
In the first stages of CCSNe, neutrinos are not in equilibrium and are not included in the EoS, but treated in transport schemes.
In later stages of CCSNe, and in (proto-) neutron stars ((P)NSs), neutrinos are trapped and weak interactions are in equilibrium.
They can thus be included in the EoS, and a lepton (or neutrino) fraction can be introduced.
For (mature) cold NSs, beta equilibrium without neutrinos is usually achieved since neutrinos become untrapped, and the electron fraction is fixed by charge neutrality together with the beta-equilibrium condition.

The determination of an EoS for compact objects is one of the main challenges in nuclear astrophysics, because of the wide range of densities, temperatures, and isospin asymmetries encountered in these astrophysical objects.
Moreover, current nuclear physics experiments cannot probe all the physical conditions found in compact stars, thus theoretical models are required to extrapolate to unknown regions. 
The set of independent thermodynamic variables for the most general EoS are usually the baryon density $n_B$, the temperature $T$, and the charge fraction, e.g. the electron fraction $Y_e$, or alternatively the charge density (see also the CompOSE manual \cite{typ2015} for an explanation of the thermodynamic variables and potentials). 
However, for cold NSs, the temperature (below 1~MeV) is lower than typical nuclear energies and the zero-temperature approximation can be adopted, thus making, together with the beta-equilibrium condition, the independent variables of the EoS reduced to the density only.
On the other hand, in CCSNe, in compact-star mergers, and in black-hole (BH) formation, the temperature can rise to a few tens or even above a hundred MeV.
Therefore, the approximate range of thermodynamic variables over which the most general EoS has to be computed is: $10^{-11} \lesssim n_B \lesssim$~few~fm$^{-3}$, $0 \leq T \lesssim 150$~MeV, and $0 < Y_e <  0.6$ (see, e.g., Figs.~1-2 in \cite{oertel2017} and \cite{oconn2011}).

In this Chapter, we aim to give an overview of the present status of the EoSs for compact-star modelling, with particular focus on the underlying many-body methods, and to discuss some of the current challenges in the field.
After a brief introduction on the nucleon-nucleon interaction in Sect.~\ref{sec:nn-inter}, we will review the theoretical many-body methods in Sect.~\ref{sec:models}, both microscopic (Sect.~\ref{sec:models-abinitio}) and phenomenological (Sect.~\ref{sec:models-phenom}).
In Sect.~\ref{sec:constraints} we will discuss the constraints on the EoS obtained in both nuclear physics experiments (Sect.~\ref{sec:constraints-nuclear}) and astrophysical observations (Sect.~\ref{sec:constraints-astro}).
We will present in Sect.~\ref{sec:eos-astro} the application of the EoSs in compact-object modelling: we will first discuss the zero-temperature NS case (Sect.~\ref{sec:eos-ns}), then we will introduce some widely used general purpose EoSs and discuss their impact in CCSNe, BH formation, and in binary mergers (Sect.~\ref{sec:eos-genpurp}). 
A brief description of the available online databases on the EoSs is given in Sect.~\ref{sec:compose}.
Finally, in Sect.~\ref{sec:future}, we will discuss some of the current challenges for the EoS modelling and in Sect.~\ref{sec:conclusions} we will draw our conclusions.

\section{Current status of many-body methods and equation of state}
\label{sec:status}

\subsection{The nucleon-nucleon interaction : a brief survey}
\label{sec:nn-inter}

The properties of the nuclear medium are strongly determined by the features of the
nucleon-nucleon (NN) interaction, in particular the presence of a hard repulsive core.
The nuclear Hamiltonian should in principle be derived from the quantum chromodynamics (QCD), but this is a very difficult task which presently cannot be realised.
There are three basic classes of bare nucleonic interactions:
\begin{itemize}
\item[-] Phenomenological interactions mediated by meson exchanges;
\item[-] Chiral expansion approach;
\item[-] Models that include explicitly the quark-gluon degrees of freedom.
\end{itemize}

In the phenomenological approaches, quark degrees of freedom are not treated explicitly but are replaced by hadrons - baryons and mesons - in which quarks are confined. 
Very refined and complete phenomenological models have been constructed for the NN interactions, e.g. the Paris potential \cite{1980PhRvC..21..861L}, the Bonn potential \cite{2001JPhG...27R..69M}, the Nijmegen potentials \cite{1977PhRvD..15.2547N,1978PhRvD..17..768N} also with hyperon-nucleon (YN) \cite{1989PhRvC..40.2226M} and hyperon-hyperon (YY) potentials \cite{1999PhRvC..59...21R}. 
Those phenomenological models have been tested using thousands of experimental data on NN scattering cross sections, from which the phase shifts in different two-body channels are extracted with high precision up to an energy of about 300 MeV in the laboratory, even if discrepancies between the results of different groups still persist \cite{2001JPhG...27R..69M}. 

The most widely known potential models are the Urbana \cite{1981NuPhA.359..331L} and Argonne potentials, the latest version called the $v18$ potential \cite{v18}. 
The structure of the NN potential is very complex and depends on many quantities characterising a two-nucleon system.
These quantities enter via operator invariants consistent with the symmetries of the strong interactions, and involve spin, isospin, and orbital angular momentum.  
The NN potential acting between a nucleon pair $ij$ is a Hermitian operator $\hat v_{ij}$ in coordinate, spin, and isospin spaces.
A sufficiently generic form of $\hat v_{ij}$ able to reproduce the abundance of NN scattering data is 

\beq
\hat v_{ij} = \sum_{u=1}^{18} v_u(r_{ij}) \hat O_{ij}^u,
\label{eq:av18}
\eeq 
where the first fourteen operators are charge-independent, i.e., invariant with respect to rotation in the isospin space:
\begin{eqnarray}
\hat O_{ij}^{u=1,...14} & =& 1,\, \mathbf{\tau}_i \cdot \mathbf{\tau}_j,\, \mathbf{\sigma}_i \cdot \mathbf{\sigma}_j, \, (\mathbf{\sigma}_i \cdot \mathbf{\sigma}_j)  (\mathbf{\tau}_i \cdot \mathbf{\tau}_j), \, \hat S_{ij} , \,  \hat S_{ij} (\mathbf{\tau}_i \cdot \mathbf{\tau}_j),  \nonumber \\
& & \hat {\mathbf{L}} \cdot \hat {\mathbf{S}} , \, \hat {\mathbf{L}} \cdot \hat {\mathbf{S}} \, (\mathbf{\tau}_i \cdot \mathbf{\tau}_j), \,\hat L^2, 
\, \hat L^2  (\mathbf{\tau}_i \cdot \mathbf{\tau}_j), \, \hat L^2  (\mathbf{\sigma}_i \cdot \mathbf{\sigma}_j), \nonumber \\
& & \hat L^2  (\mathbf{\sigma}_i \cdot \mathbf{\sigma}_j)  (\mathbf{\tau}_i \cdot \mathbf{\tau}_j), \,  (\hat {\mathbf{L}} \cdot \hat {\mathbf{S}})^2, \, (\hat {\mathbf{L}} \cdot \hat {\mathbf{S}})^2 (\mathbf{\tau}_i \cdot \mathbf{\tau}_j).
\label{eq:av18_op}
\end{eqnarray}
The notation $\mathbf{r}_{ij} = \mathbf{r}_i - \mathbf{r}_j$ indicates the relative position vector, whereas $\mathbf{\sigma}_i$ and $\mathbf{\sigma}_j$ are spins (in units of $\hbar /2$), and $\mathbf{\tau}_i$ and $\mathbf{\tau}_j$ are isospins (in units of $\hbar /2$). 
The relative momentum is denoted by $\hat {\mathbf{p}}_{ij} =\hat {\mathbf{p}_i}- \hat {\mathbf{p}_j}$; $\hat {\mathbf{L}} = {\mathbf{r}_{ij}} \times {\hat{\mathbf{p}}}_{ij} $ is the total orbital angular momentum, and $\hat {L}^2 $ its square in the centre-of-mass system.  The spin-orbit coupling enters via $\hat {\mathbf{L}} \cdot \hat {\mathbf{S}}$,
being $\hat {\mathbf{S}} = (\mathbf{\sigma}_i + \mathbf{\sigma}_j )/2$ the total spin (in units of $\hbar$). 
Analogously, we define the total isospin $\hat {\mathbf{T}} = (\mathbf{\tau}_i + \mathbf{\tau}_j )/2$.  
The tensor coupling enters via the tensor operator
\beq
\hat S_{ij} =3(\mathbf{\sigma}_i \cdot \mathbf{n}_{ij}) (\mathbf{\sigma}_j \cdot \mathbf{n}_{ij}) - \mathbf{\sigma}_i \cdot \mathbf{\sigma}_j \ ,
\label{eq:tens}
\eeq
where $\mathbf{n}_{ij} = \mathbf{r}_{ij}/r_{ij}$. 
Both the spin-orbit and tensor couplings are necessary for explaining experimental data.
The terms with $\hat O_{ij}^{u=15,...18}$ are small and break charge independence, and they correspond to $v_{np} (T = 1) = v_{nn} = v_{pp}$, while the charge symmetry implies only that $v_{nn} = v_{pp}$. Modern fits to very precise nucleon scattering data indicate the existence of charge-independence breaking. However, the effect of such forces on the energy of nucleonic matter is much smaller than the uncertainties of many-body calculations and therefore can be neglected while constructing the EoS. 

A different approach to the study of the NN interaction is the one based on quark and gluon degrees of freedom, thus connecting the low energy nuclear physics phenomena with the underlying QCD structure of the nucleons.
This is quite difficult because the whole hadron sector is in the non-perturbative regime, due to confinement. 
A possible strategy is based on the systematic use of the symmetries embodied in the hadronic QCD structure. 
The main symmetry which is explicitly broken in the confined matter is the chiral symmetry, since the bare $u$ and $d$ quark masses in non-strange matter are just a few MeV. 
According to the general Goldstone theorem, this results in the physical mass of the pion, which suggests to treat the pion degrees of freedom explicitly and to describe the short range part by structureless contact terms. 
Along this line, Weinberg \cite{1990PhLB..251..288W,1991NuPhB.363....3W} proposed a scheme for including in the interaction a series of operators which reflect the partially broken chiral symmetry of QCD. 
The strength parameters associated to each operator are then determined by fitting the NN phase shifts, the properties of deuteron and of few-body nuclear systems.
The method is then implemented in the framework of the Effective Field Theory (EFT), i.e. by ordering the terms according to their dependence on the physical parameter $q/m$, where $m$ is the nucleon mass and $q$ a generic momentum that appears in the Feynman diagram for the considered process. 
This parameter is assumed to be small and each term is dependent on a given power of this parameter thus fixing its relevance. 
In this way a hierarchy of the different terms of the forces is established. 
In particular, the pion exchange term is treated explicitly and is considered the lowest order (LO) term of the expansion. 
Moreover, it is found that the three-body forces (TBFs) so introduced are of higher order than the simplest two-body forces and they are treated on an equal footing. 
They arise first at next-to-next leading order (N$^2$LO) and, as a consequence, because of the hierarchy intrinsic in the chiral expansion, TBFs are expected to be smaller than two-body forces, at least within the range of validity of the expansion, whereas four-body forces appear only at next-to-next-to-next leading order (N$^3$LO) level, and so on. 
It has to be stressed that in the TBFs the same couplings that fix the two-body forces have to be used and, in general, only a few additional parameters must be introduced as the order increases. 
Therefore the TBFs are automatically consistent with the two-body forces, and so on for the higher order many-nucleon forces. 
At present the nucleonic interaction has been calculated up to N$^4$LO \cite{hu2017}.
An exhaustive list of higher order diagrams up to N$^3$LO can be found in review papers \cite{2005NuPhA.751..149M, 2009RvMP...81.1773E}.
This Chiral Perturbation Expansion (ChPE) can be used to construct NN interactions that are of reasonably good quality in reproducing the two-body data \cite{Machleidt_PLB,Holt:2009ty}. 
The assumption of a small $q/m$ parameter in principle restricts the applications of these forces to not too large momenta, and therefore to a not too large density of nuclear matter. 
It turns out that the safe maximum density is around the saturation value, $n_0$. 
This method has been refined along the years and many applications can be found in the literature. 

Another approach inspired by the QCD theory of strong interaction has been developed in \cite{Fujiwara:2006yh,Oka:1986fr,Oka:1980ax,Shimizu:1989ye,Valcarce:2005em}. 
In this approach, based on the resonating-group method (RGM),  the quark degree of freedom is explicitly introduced and the NN interaction is constructed from gluon and meson exchange between quarks, the latters being confined inside the nucleons. 
The resulting interaction is highly non-local due to the RGM formalism and contains a natural cut-off in momentum.
The most recent model, named fss2 \cite{Fujiwara:2001jg,Fujiwara:2006yh}, reproduces closely the experimental phase shifts, and fairly well the  data on the few-body systems, e.g. the triton binding energy is reproduced within 300 keV. 
Recently, it has been shown that the fss2 interaction is able to reproduce correctly the nuclear matter saturation point without the TBF contribution \cite{Baldo:2014rda}.

Recently, a further possibility of constructing the NN interaction based on lattice QCD has been explored, see \cite{2012PTEPaA105A,2011PrPNP..66....1B} for a review. 
This tool, from which one should be able in principle to calculate the hadron properties directly from the QCD Lagrangian, is extremely expensive from the numerical pont of view and current simulations can be performed only with large quark masses. 
In fact, an accurate simulation has to be made on a fine grid spacing and large volumes, thus requiring high performance computers.
Hopefully in the next few years high precision calculations will be possible, especially for those channels where scarce experimental data are available, e.g. the nucleon-hyperon interaction.
  
A further class of NN interactions is based on renormalization group (RG) methods (see, e.g., \cite{Machleidt_PLB, RG} for a complete review). 
The main effect of the hard core in the NN interaction is to produce scattering to high momenta of the interacting particles. 
A possible way to soften the hard core from the beginning is by integrating out all the momenta larger than a certain cut-off $\Lambda$ and ``renormalize'' the interaction to an effective interaction V$_{low}$ in such a way that it is equivalent to the original interaction for momenta $q \, <\, \Lambda$. 
The V$_{low}$ interaction turns out to be much softer, since no high momentum components are present and, as a consequence, three- and many-body forces emerge automatically from a pure two-body force.  The short range repulsion is replaced by the non local structure of the interaction. 
The cut-off $\Lambda$ is taken above $300$ MeV in the laboratory, corresponding to relative momentum $q \, \approx\,2.1$~fm$^{-1}$, that is the largest energy where the experimental data are established. 
The fact that V$_{low}$ is soft has the advantage to be much more manageable than a hard core interaction, in particular it can be used in perturbation expansion and in nuclear structure calculations in a more efficient way \cite{RG,2013RPPh...76l6301F}.

\subsection{Theoretical many-body methods}
\label{sec:models}

The theoretical description of matter in extreme conditions is a very challenging task. 
Moreover, current nuclear physics experiments cannot probe all the physical conditions encountered in compact stars.
Therefore, theoretical models are required to extrapolate to unknown regions.
The undertaken theoretical approaches also depend on the relevant degrees of freedom of the problem, from nuclei and nucleons at lower densities and temperature, to additional particles, such as hyperons and quarks, at high densities and temperature.
The current theoretical many-body approaches to describe a nuclear system can be divided into two main categories: 
\begin{enumerate}
\item {\bf Ab-initio (microscopic) approaches}, that start from ``realistic'' two-body interactions fitted to experimental NN scattering data and to the properties of bound few-nucleon systems.
Examples of these kinds of models are Green's function methods, (Dirac-)Brueckner Hartree-Fock, variational, coupled cluster, and Monte Carlo methods.
Despite the tremendous progress that has been done in the last years, these methods cannot yet be applied to large finite nuclear systems. 
Nevertheless, recent developments allow ab-initio methods to reach medium to ``heavy'' nuclei, see e.g. \cite{cbn2013,  sbd2013, binder2014, cbn2015}.
Therefore, in the description of dense matter, the ab-initio models are usually restricted to homogeneous matter; thus, they are not applied to describe clustered matter (like in SN cores or NS crusts).
\item {\bf Phenomenological approaches}, that rely on effective interactions which depend on a certain number of parameters fitted to reproduce properties of finite nuclei and nuclear matter.
This class of methods are widely used in nuclear structure and astrophysical applications. 
Among them, there are self-consistent mean-field models and shell-model approaches.
In astrophysics, the latters have been employed, for example, to study electron-capture rates on nuclei relevant for SN simulations (see e.g.~\cite{lanmar2014} and references therein).
Alternatively, models based on self-consistent mean-field approaches are widely used, in particular, to build EoSs of dense matter.
These methods, based on the nuclear energy-density functional (EDF) theory, can be either non-relativistic (e.g. using Skyrme or Gogny interactions) or relativistic (based on an effective Lagrangian with baryon and meson fields).
A more macroscopic approach to treat the many-body system is the (finite-range) liquid-drop model, which parameterizes the energy of the system in terms of global properties such as volume energy, asymmetry energy, surface energy, etc. and whose parameters are fitted phenomenologically.
The liquid-drop model usually describes well the trend of nuclear binding energies and has been largely applied to construct EoSs for compact stars.
\end{enumerate}

In the following, we will not aim at giving a complete review on the different theoretical many-body approaches (see, e.g.,~\cite{ringschuck, Muether00, bender2003, balbur2012, duguet2014, 2015RvMP...87.1067C}), but we will give an overview of the two kinds of approaches, focusing on the latest advances.

\subsubsection{Ab-initio approaches}
\label{sec:models-abinitio}

A microscopic many-body method is characterised mainly by two basic elements: the realistic bare interaction among nucleons and the many-body scheme followed in the calculation of the EoS. 
The many-body methods can be enumerated as follows:

\begin{itemize}
\item[-] The Bethe-Brueckner-Goldstone (BBG) diagrammatic method and the corresponding hole-line expansion,
\item[-] The relativistic Dirac-Brueckner Hartree-Fock (DBHF) approach, 
\item[-] The variational method,
\item[-] The coupled cluster expansion,
\item[-] The self-consistent Green's function (SCGF),
\item[-] The renormalization group (RG) method,
\item [-] Different methods based on Monte Carlo (MC) techniques.
\end{itemize}

A brief survey of all those methods is given below. 
For further details the reader is left to the quoted references. 

\begin{itemize}

\item{\it{The Bethe-Brueckner-Goldstone expansion.}}\label{BBG}

The BBG many-body theory is based on the re-summation of the perturbation expansion of the ground-state energy of nuclear matter \cite{book,jpg}. 
The original bare NN interaction is systematically replaced by an effective interaction that describes the in-medium scattering processes, the so-called $G$-matrix, that takes into account the effect of the Pauli principle on the scattered particles, and the in-medium potential $U(k)$ felt by each nucleon, $k$ being the momentum.
 The corresponding integral equation for the $G$-matrix can be written as
 \begin{eqnarray}
 \bra k_1 k_2 \vert G(\omega) \vert k_3 k_4 \ket &=&
 \bra k_1 k_2 \vert v \vert k_3 k_4 \ket 
 + \sum_{k'_3 k'_4} \bra k_1 k_2 \vert v \vert k'_3 k'_4 \ket \nonumber \\
 & \times & \frac{\left(1 - \Theta_F(k'_3)\right) \left(1 - \Theta_F(k'_4)\right)}
   {\omega - e_{k'_3} + e_{k'_4} + i \eta}
  \, \bra k'_3 k'_4 \vert G(\omega) \vert k_3 k_4 \ket \ \ ,
 \label{eq:bruin} 
\end{eqnarray}
\noindent where $v$ is the bare NN interaction, $\omega$ is the starting energy, the two factors $(1 - \Theta_F(k))$ force the intermediate momenta to be above the Fermi momentum (``particle states''), the single-particle energy being $e_k \, =\, \hbar^2 k^2/ 2m \, +\, U(k)$, with $m$ the particle mass, and the summation includes spin-isospin variables.
The main feature of the $G$-matrix is that it is defined even for bare interactions with an infinite hard core, thus making the perturbation expansion more manageable. 
The introduction and choice of the in-medium single-particle potential are essential to make the re-summed expansion convergent. 
The resulting nuclear EoS can be calculated with good accuracy in the Brueckner two hole-line  approximation with the continuous choice for the single-particle potential, the results in this scheme being quite close to the calculations which include also the three hole-line contribution \cite{Song:1998zz}.  

\par One of the well known results of all non-relativistic many-body approaches is the need of introducing TBFs in order to reproduce correctly the saturation point in symmetric nuclear matter. 
For this purpose, TBFs are reduced to a density dependent two-body force by averaging over the generalised coordinates (position, spin, and isospin) of the third particle, assuming that the probability of having two particles at a given distance is reduced according to the two-body correlation function.
In the BBG calculations for nuclear matter, a phenomenological approach to the TBF was adopted, based on the so-called Urbana model for finite nuclei, which consists of an attractive two-pion exchange contribution between two nucleons via the excitation of a third nucleon, e.g. a $\Delta$-baryon \cite{1957PThPh..17..360F}, supplemented by a parameterized repulsive part \cite{UIX,2008AIPC.1011..143P,2001PhRvC..64a4001P,1995PhRvL..74.4396P}, adjusted to the properties of light nuclei. 
In the nuclear matter case, the two parameters contained in the Urbana TBF~\cite{Baldo:1997ag,Zhou:2004br, Li:2006gr} were accurately tuned in order to get an optimal nuclear matter saturation point. 
In symmetric nuclear matter, this TBF produces a shift in the binding energy of about $+1$~MeV and of $-0.01$~fm$^{-3}$ in density. 
The problem of such a procedure is that the TBF is dependent on the two-body force.
The connection between two-body and TBFs within the meson-nucleon theory of nuclear interaction is extensively discussed and developed in \cite{Zuo:2002sfa,zuo_2002}. 
At present the theoretical status of microscopically derived TBFs is still quite rudimentary; however, a tentative approach has been proposed using the same meson-exchange parameters as the underlying NN potential. 
Results have been obtained with the Argonne $v$18~\cite{v18}, the Bonn B \cite{Bonn}, and the Nijmegen 93 potentials \cite{li2,Li1}.
Alternatively, latest nuclear matter calculations \cite{2015PhRvC..91f4001L} used a new class of chiral inspired TBF, showing that the considered TBF models are not able to reproduce simultaneously the correct saturation point and the properties of three- and four-nucleon systems.

Recently, it has been shown that the role of TBF is greatly reduced if the NN potential is based on a realistic constituent quark model~\cite{Baldo:2014rda} which can explain at the same time few-nucleon systems and nuclear matter, including the observational data on NSs and the experimental data on heavy-ion collisions (HICs) \cite{Ken}. 
An extensive comparison among several EoSs obtained using different  two-body and TBFs is illustrated afterwards.

\item{\it{The Dirac-Brueckner Hartree-Fock approach.}}\label{DBHF}

The relativistic approach is the framework on which the nuclear EoS should be ultimately based.
The best relativistic treatment developed so far is the Dirac-Brueckner approach, about which excellent review papers can be found in the literature (see, e.g.,~\cite{Machleidt1989}).
In the relativistic context, the only two-body forces that have been used are the ones based on meson exchange models. 
The DBHF method has been developed in analogy with the non-relativistic case, where the two-body correlations are described by introducing the in-medium relativistic $G$-matrix. 
This is a difficult task, and in general one keeps the interaction as instantaneous (static limit) and a reduction to a three-dimensional formulation from a four-dimensional one.
The main relativistic effect is due to the use of the spinor formalism which has been shown \cite{Brown87} to be equivalent to introducing a particular TBF, the so-called Z-diagram.
This TBF turns out to be repulsive and consequently produces a saturating effect. 
In fact the DBHF gives a better saturation point than the BHF. 
In this way, a definite link between DBHF and BHF + TBF is established. Indeed, including in BHF only these particular TBFs, one gets results close to DBHF calculations, see e.g.~\cite{Li:2006gr}.
Generally speaking, the EoS calculated within the DBHF method turns out to be stiffer above saturation than the ones calculated from the BHF + TBF method.
Currently, some features of this method are still controversial and the results depend strongly on the method used to determine the covariant structure of the in-medium $G$-matrix.

\item{\it {The variational method.}}\label{Var}

In the variational method one assumes that the ground-state trial wave function $\Psi$ can be written as 
\beq
     \Psi_{trial} (r_1,r_2,......) \, =\, \prod_{i<j} f(r_{ij}) \Phi(r_1,r_2,.....)
     \,\,\,\, ,
\le{trial} \eeq \noindent 
where $\Phi$ is the unperturbed ground-state wave function, properly antisymmetrised, and the product runs over all possible distinct pairs of particles. 
The correlation factors $f$ are  determined by the Ritz-Raleigh variational principle, i.e. by imposing that the mean value of the Hamiltonian gets a minimum 
\beq
   \frac{\delta}{ \delta f} \frac{\bra \Psi_{trial} \vert H \vert \Psi_{trial} \ket } 
   {\bra \Psi_{trial} \vert \Psi_{trial} \ket} \,= \, 0 \,\,\, .
\le{euler} \eeq \noindent
In principle this is a functional equation for  $f$ and it is intended to transform the uncorrelated wave function $\Phi(r_1,r_2,.....)$ to the correlated one, and can be written explicitly in a closed form only if additional suitable approximations are introduced. 
Once the trial wave function is determined, all the expectation values of other operators can be calculated. 
Therefore the main task in the variational method is to find a suitable ansatz for the correlation factors $f$.
Several different methods exist for the calculation of $f$, e.g. in the nuclear context  the Fermi-Hyper-Netted-Chain (FHNC) \cite{1975NCimA..25..593F,1979RvMP...51..821P} calculations have been proved to be efficient.

For nuclear matter at low densities, two-body correlations play an essential role, and this justifies the assumption that $f$ is actually a two-body operator $\hat{F}_{ij}$.
Generally one assumes that $\hat{F}$ can be expanded in the same spin-isospin, spin-orbit, and tensor operators appearing in the NN interaction \cite{1998LNP...510..119F,2015RvMP...87.1067C}.  
Due to the formal structure of the Argonne NN forces, most variational calculations have been performed with this class of NN interactions, often supplemented by the Urbana TBFs.
Many excellent review papers exist in the literature on the variational method and its extensive use for
the determination of nuclear matter EoS, e.g.~\cite{1979RvMP...51..821P, 2002immq.book..121N}. 
The best known and most used variational nuclear matter EoS is the Akmal-Pandharipande-Ravenhall (APR)~\cite{apr}. 
A detailed discussion on the connection between variational method and BBG expansion can be found in~\cite{jpg}.

Other methods based on the variational principle are widely used in nuclear physics to evaluate expectation values. Among those, we mention the coupled-cluster theory, proposed in \cite{1958NucPh...7..421C,1960NucPh..17..477C}, in which the correlation operator is represented in terms of the cluster operator.
The method has been proved to be successful in recent nuclear matter calculations with chiral NN interactions~\cite{2007PhRvC..76c4302H,2014PhRvC..89a4319H} and also in nuclear structure calculations~\cite{2012PhRvL.109c2502H, 2010PhRvC..82c4330H}.
The variational Monte Carlo (VMC) approach is also widely used in nuclear physics to evaluate expectation values. 
Several calculations have been performed for light nuclei, including two and three-body correlations \cite{2014PhRvC..89b4305W}, but the EoS of homogeneous nuclear matter is hard to obtain, due to the increasingly large computational effort with the number of nucleons (see~\cite{2002immq.book..121N,2015RvMP...87.1067C} for complete reviews).

\item{\it {Chiral effective field theory ($\chi$EFT) approach.}}\label{EFT}

High-precision nuclear potentials based on chiral perturbation theory (ChPT) \cite{2003PhRvC..68d1001E,2011PhR...503....1M} are nowadays widely employed to link QCD, the fundamental theory of strong interactions, to nuclear many-body phenomena. 
In particular, for nuclear matter, many-nucleon forces are of course relevant. 
In this case another scale appears, $k_F/m$, $k_F$ being the Fermi momentum, which is of the same order of the pion mass $m_\pi$ at saturation and it is smaller than a typical hadron scale. 
In the chiral limit it is then natural to expand in $k_F/m$, and this expansion can be obtained from the vacuum ChPT expansion \cite{2002NuPhA.697..255K}. 
For nuclear matter the correction thus obtained with respect to the vacuum diagrams gives a direct contribution to the EoS of nuclear matter, and this correction is clearly proportional to a power of $k_F/m$. 
Also in this case a cut-off must be introduced, and its tuning allows to obtain a saturation point and compressibility in fair agreement with phenomenology.
Along the same lines more sophisticated expansions can be developed, including a power counting modified for finite density systems, where the small scale is fixed by both $k_F$ and $m_\pi/m$. 
The results thus obtained are in good agreement with the most advanced non-relativistic many-body calculations \cite{2011AnPhy.326..241L}.
A different approach can be developed, where the many-nucleon interactions built in vacuum are directly used in nuclear matter calculations. 
In this case the ChPT is used in conjunction with the EFT scheme.
In recent years, $\chi$EFT has been used for studying nuclear matter within various theoretical frameworks like many-body perturbation theory \cite{2011PhRvC..83c1301H, 2014PhRvC..89f4009W, 2014PhRvC..89d4321C, 2016PhRvC..93e4314D}, SCGF framework \cite{2013PhRvC..88d4302C}, in-medium chiral perturbation theory \cite{2013PrPNP..73...35H}, the BHF approach \cite{2013PhRvC..88f4005K,2012PhRvC..85f4002L}, and quantum Monte Carlo methods \cite{2013PhRvL.111c2501G, 2014PhRvL.112v1103R, 2016PhRvL.116f2501L}.
Several reliable calculations have been performed up to twice the saturation density $n_0$, beyond which uncertainties were estimated by analysing the order-by-order convergence in the chiral expansion and the many-body perturbation theory \cite{2014PhRvC..89d4321C,2017PhRvC..95c4326H}.
Variations in the resolution scale \cite{2005NuPhA.763...59B} and low-energy constants appearing in the two-nucleon and three-nucleon forces were sistematically explored \cite{2010PhRvC..82a4314H}.
It has been found that the theoretical uncertainty band grows rapidly with the density beyond $n_0$, due to the missing third-order terms at low densities and higher-order contributions in the chiral expansion. 
This has consequences not only for the EoS, but also for the symmetry energy at saturation density, $S_0$, and the slope parameter $L$, as discussed in \cite{2016PrPNP..91..203B}.

\item{\it {Self-consistent Green's function.}}\label{SCGF}
\label{sec:scgf}

Another way to approach the many-body problem is through the many-body Green's functions formalism \cite{dickhoff08}. 
In this approach one performs a diagrammatic analysis of the many-body propagators in terms of free one-body Green's functions and two-body interactions. 
The perturbative expansion results in an infinite series of diagrams, among which one has to choose those which are relevant for the considered physical problem. 
Depending on the approximation, one can either choose a given number of diagrams or sum an infinite series of them, in analogy with the BHF approach. In the description of nuclear matter, the method is conventionally applied at the ladder approximation level, which encompasses at once particle-particle and hole-hole propagation, and this represents the main difference with respect to the $G$-matrix, where only particle-particle propagators are included.
At a formal level, the comparison between the BHF and the SCGF approaches is not straightforward. 
Even though both approaches arise from a diagrammatic expansion, the infinite subsets of diagrams considered in the two approaches are not the same, and the summation procedures are also somewhat different. 
Whereas the BHF formalism in the continuous choice can be derived from the ladder SCGF formalism after a series of approximations, this is not the case for the full BBG expansion. 
In principle, if both BBG and SCGF were carried out to all orders, they should yield identical results. 
BBG theory, however, is an expansion in powers of density (or hole-lines), and the three-hole line results seem to indicate that it converges quickly. 
The error in the SCGF expansion is more difficult to quantify, as one cannot directly compute (or even estimate) which diagrams have to be included in the expansion. 
Reviews on the applications of the method to nuclear problems can be found in \cite{Muether00, Dickhoff04}. 

Also for the SCGF method the inclusion of TBFs is essential. 
So far TBFs were not included in the ladder approximation, however a method has been developed recently in~\cite{2013PhRvC..88e4326C}, and applied to symmetric nuclear matter using chiral nuclear interactions. 
TBFs are included via effective one-body and two-body interactions, and are found to improve substantially the saturation point~\cite{2013PhRvC..88d4302C}.

One has to notice that because of the well-known Cooper instability \cite{1956PhRv..104.1189C}, through which a
fermionic many-body system with an attractive interaction tends to form pairs at the Fermi surface, low-temperature nuclear matter is unstable with respect to the formation of a superfluid or superconducting state. 
The Cooper instability shows up as a pole in the $T$-matrix when the temperature falls below the critical temperature for the transition to the superfluid/superconducting state. 
Therefore current calculations are often performed at temperatures above the critical temperature and extrapolated to zero temperature, see \cite{Frick04} for details.

\item{\it {Quantum Monte Carlo methods.}}\label{QMC}

Quantum Monte Carlo (QMC) methods are very successful in describing the ground state of fermionic systems, like liquid $^3$He, or bosons, like atomic liquid $^4$He.  
Modern computer technology has allowed the extension of the QMC method to nuclear systems, which have more complicated interactions and correlation structures.
The mostly used versions are the auxiliary field diffusion Monte Carlo (AFDMC) \cite{AFDMC} and the Green's function Monte Carlo (GFMC) \cite{GFMC} methods, which differ in the treatment of the spin and isospin degrees of freedom. 
It has to be noticed that the computing time increases exponentially with the number of particles, which limits the number of nucleons considered by GFMC up to 16 neutrons. 
The largest nucleus considered is $^{12}$C. 
The AFDMC strategy allows to efficiently sample spin-isospin correlations in systems with a sufficient number of nucleons (N = 114). 
A recent comparison has demonstrated that both methods give very close results for neutron drops with N $\leq$ 16 \cite{2011PhRvL.106a2501G}.
However, the accuracy of the different QMC versions is limited by the fermion sign problem \cite{1987TCP....36..125S},  for which different approximations are adopted \cite{2012PTEP.2012aA209C,2015RvMP...87.1067C}. 
This seriously limits the potentiality of the QMC approach. 

In spite of its recent progress, it is not yet possible to perform GFMC and AFDMC calculations with the Argonne $v18$ potential, mainly due to technical problems associated with the spin-orbit structure of the interaction and the trial wave function, which induce very large statistical errors. 
In order to overcome this problem, the full operatorial structure of current high-quality NN potentials has been simplified and more manageable NN potentials have been developed containing less operators with readjusted parameters. 
In particular, we mention the V8', V6', and V4' potentials \cite{1997PhRvC..56.1720P,2002PhRvL..89r2501W}, eventually supplemented with the Urbana TBFs. 
Recently, a local chiral potential has been developed~\cite{2013PhRvL.111c2501G} which is well suited for QMC techniques.

\end{itemize}

\paragraph{\bf Finite-temperature equation of state}

In the latest stage of the SN collapse the EoS of asymmetric nuclear matter at finite temperature plays a major role in determining the final evolution. 
Microscopic calculations of the nuclear EoS at finite temperature are quite few. 
The variational calculation by Friedman and Pandharipande \cite{fp_1981} was one of the first few
semi-microscopic investigations. 
In the resulting EoS for symmetric nuclear matter, one recognizes the familiar Van der Waals shape, which entails a liquid-gas phase transition, with a definite critical temperature $T_c$, i.e. the temperature at which the minimum in the Van der Waals isotherm disappears. 
In the Friedman and Pandharipande work, the critical temperature turns out to be around $T_c = 18 - 20$ MeV.
The values of the critical temperature, however, depend on the theoretical scheme, as well as on the particular NN interaction adopted. 
In particular, non-relativistic Brueckner-like calculations at finite temperature \cite{1999PhRvC..59..682B}, where the formalism by Bloch and De Dominicis (BD) \cite{1958NucPh...7..459B,1959NucPh..10..181B,1959NucPh..10..509B} was followed, confirmed the Friedman and Pandharipande findings with very similar values of $T_c$.
The main difficulty in this approach is the lack of thermodynamic consistency. 
In fact the thermodynamic relation $P \,=  -\mathcal{F} \,+\, \mu n$, 
which connects the pressure $P$ with the free energy density $\mathcal{F}$, the chemical potential $\mu$, and the number density $n$ and usually referred to as the Hughenoltz-Van Hove theorem, is not satisfied. 
In other words, the pressure calculated in such a way does not coincide with the pressure calculated from $P \,=\, - \Omega/V$ ($\Omega$ being the grand potential and $V$ the volume). 
In \cite{1999PhRvC..59..682B}, a procedure was proposed in order to overcome this problem: the pressure is calculated from the derivative of the free energy per particle so that the Hughenoltz-Van Hove theorem is automatically satisfied. 
The difficulty is that the chemical potential determined by fixing the density in the Fermi distribution is not strictly the one extracted from the derivative of $\mathcal{F}$, as it should be. 
In any case, the procedure looks most reliable within the Brueckner scheme (see \cite{1999PhRvC..59..682B} for details).
For completeness, we remind the reader that the Brueckner approximation, both at zero and finite temperature, violates the Hugenoltz-Van Hove theorem.
On the contrary, the Hughenoltz-Van Hove theorem is strictly fulfilled within the SCGF method \cite{1962PhRv..127.1391B,2006PhRvC..74e4317R,2008PhRvC..78d4314R}.
The results at the two-body correlation level, when only two-body forces are used, in some cases are similar to the Brueckner ones, in some others they differ appreciably according to the forces used. 
The main difference with the Brueckner scheme is the introduction in the ladder summation of the hole-hole propagation, which gives a repulsive contribution. 
As a result, the critical temperature in the SCGF approach with Argonne $v_{18}$ potential is found to be about $T_c \approx 11.6$~MeV, whereas in the BHF approach $T_c \approx 18.1$~MeV~\cite{2008PhRvC..78d4314R}, depending on the adopted NN interaction.
As far as the DBHF is concerned, it turns out that the critical temperature within this scheme is definitely smaller  than in the non-relativistic scheme, about 10~MeV against 18-20~MeV \cite{1986PhRvL..56.1237T, 1987PhR...149..207H,1998PhRvC..57.3484H}.
This cannot be due to relativistic effects, since the critical density is about $1/3$ of the saturation density, but to a different behaviour of the Dirac-Brueckner EoS at low density. 
This point remains to be clarified.
Indeed, there are experimental data from heavy-ion reactions that point towards a value of $T_c > 15$~MeV \cite{2008PrPNP..61..551B, karna2008}.

\paragraph{\bf Results and discussion}
\label{sec:Results}

We will discuss here the results obtained with some of the widely used many-body methods illustrated above. 
The simplest constraint that has to be considered is the reproduction of the phenomenological saturation point; we will see that this condition is not trivially fulfilled. 
Other constraints will be analysed afterwards.  

\begin{figure} [!t]
\includegraphics[scale=0.52]{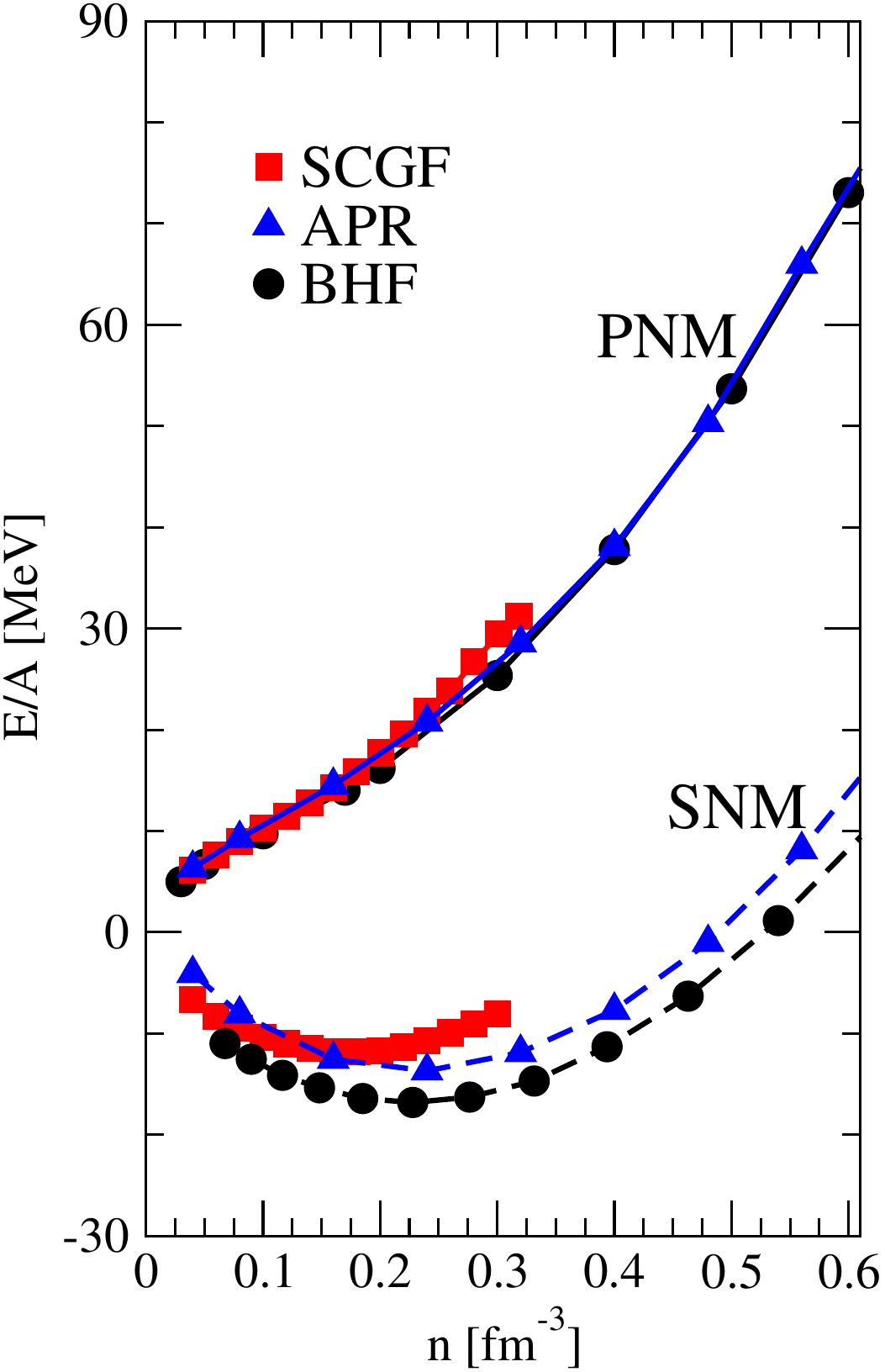}
\includegraphics[scale=0.27]{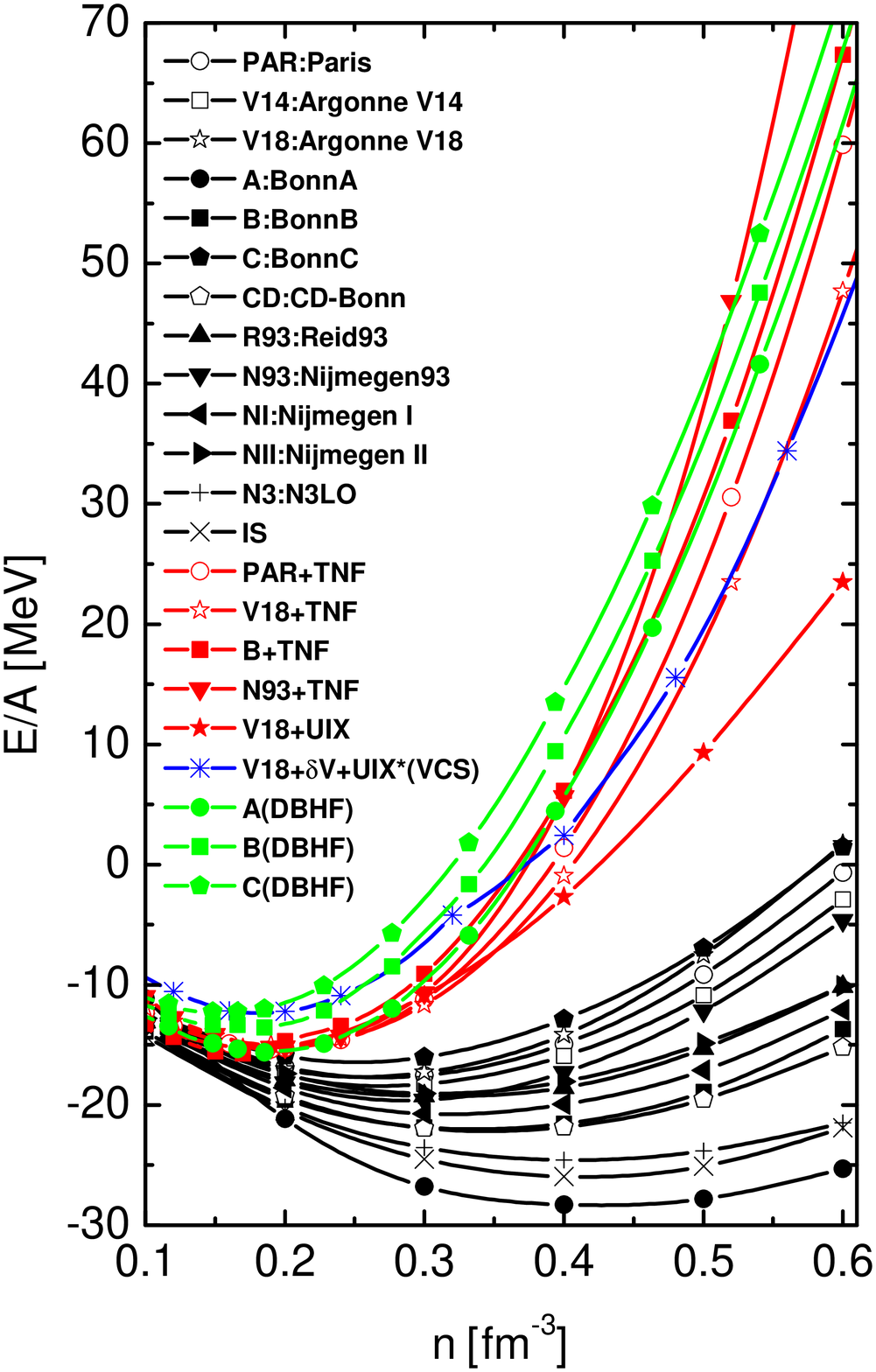}
\caption{Left panel: Symmetric and pure neutron matter EoS from BHF (black circles), SCGF (red squares), and APR (blue triangles) schemes including only two-body forces. Right panel: Energy per nucleon of symmetric nuclear matter obtained with different NN and TBF interactions, for different theoretical approaches; courtesy of H.-J. Schulze. See text for details.}
\label{fig:EOS1}
\end{figure}

We begin by discussing a comparison between the BHF, SCGF, and APR EoS with only two-body forces. 
Results are displayed for the binding energy per nucleon, $E/A$, of symmetric nuclear matter (SNM) and pure neutron matter (PNM) in Fig.~\ref{fig:EOS1}, left panel, where the Argonne $v18$ NN potential is adopted. 
We notice a substantial agreement between all methods for PNM calculations, whereas for SNM some differences show up.  
It is well known that the discrepancies between SCGF and BHF result in an overall repulsive effect in the binding energy \cite{2003PhRvL..90o2501D}, which is mainly due to the inclusion in the SCGF expansion of the hole-hole propagation. 
Those effects are quite sizeable in SNM. 
For instance, the saturation point shifts from $n_0=0.25$~fm$^{-3}$, $E(n_0)/A=-16.8$~MeV for BHF to $n_0=0.17$~fm$^{-3}$, $E(n_0)/A=-11.9$~MeV for SCGF. 
While the shift seems to go towards the right saturation density, the value of the SCGF saturation energy is quite high.

In the right panel of Fig.~\ref{fig:EOS1}, we display the energy per particle in SNM obtained with a set of NN potentials and with different TBFs (TNF in the legend).
On the standard BHF level (black curves) one obtains in general too strong binding, varying between the results with the Paris \cite{1980PhRvC..21..861L}, $v18$ \cite{v18}, and Bonn C potentials \cite{Machleidt1989,Bonn} (less binding), and those with the Bonn A~\cite{2001JPhG...27R..69M}, N$^3$LO \cite{Machleidt_PLB,2003PhRvC..68d1001E}, and IS \cite{2003PhRvC..67f4005D} potentials (very strong binding). 
Including TBFs, with the Paris, Bonn B, $v18$, and Njimegen 93 \cite{1994PhRvC..49.2950S} potentials, adds considerable repulsion and yields results slightly less repulsive than the DBHF ones with the Bonn potentials (green curves). 
This is not surprising, because it is well known that the major effect of the DBHF approach amounts to include the TBF corresponding to nucleon-antinucleon excitation by 2$\sigma$  exchange within the BHF calculation. 
In those BHF calculations microscopic TBFs have been included and those turn out to be more repulsive at high density than the phenomenological TBF, i.e. the one derived from the Urbana UIX model (full red symbols). 
This is a clear sign of uncertainty in the role of TBF at large density.
The blue curve with asterisks represents the results of the APR EoS obtained with the Urbana UIX TBF, which was adjusted to reproduce the saturation point, by varying mainly one parameter, as it has been done in the BHF approach. 
We notice that the effect of the TBF is quite moderate around saturation ($\delta n = -0.01$~fm$^{-3}$, 
$\delta E = + 1$~MeV), but they are essential to get the correct saturation point. 

\begin{figure} [!t]
\includegraphics[scale=0.46]{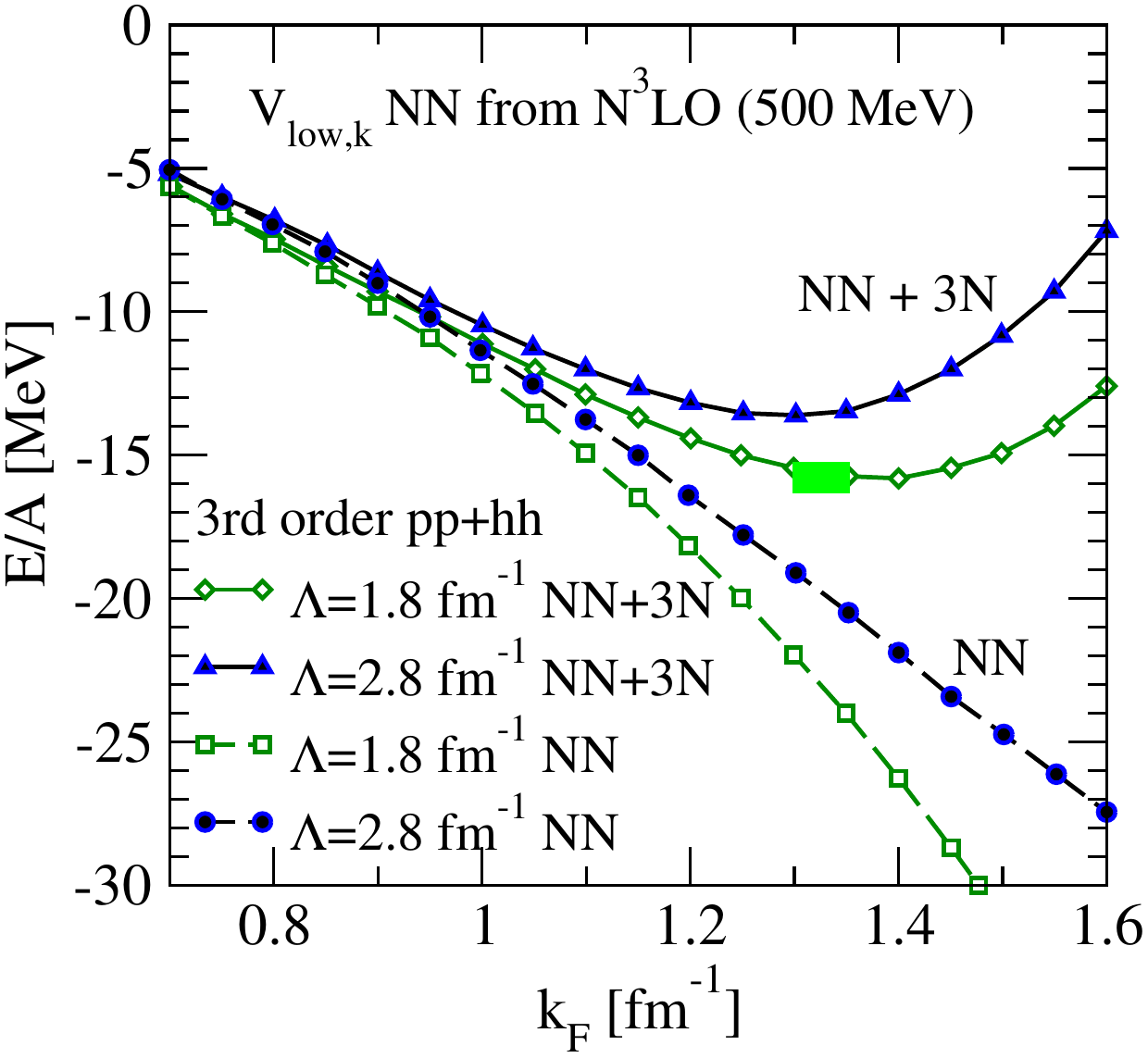}
\includegraphics[scale=0.46]{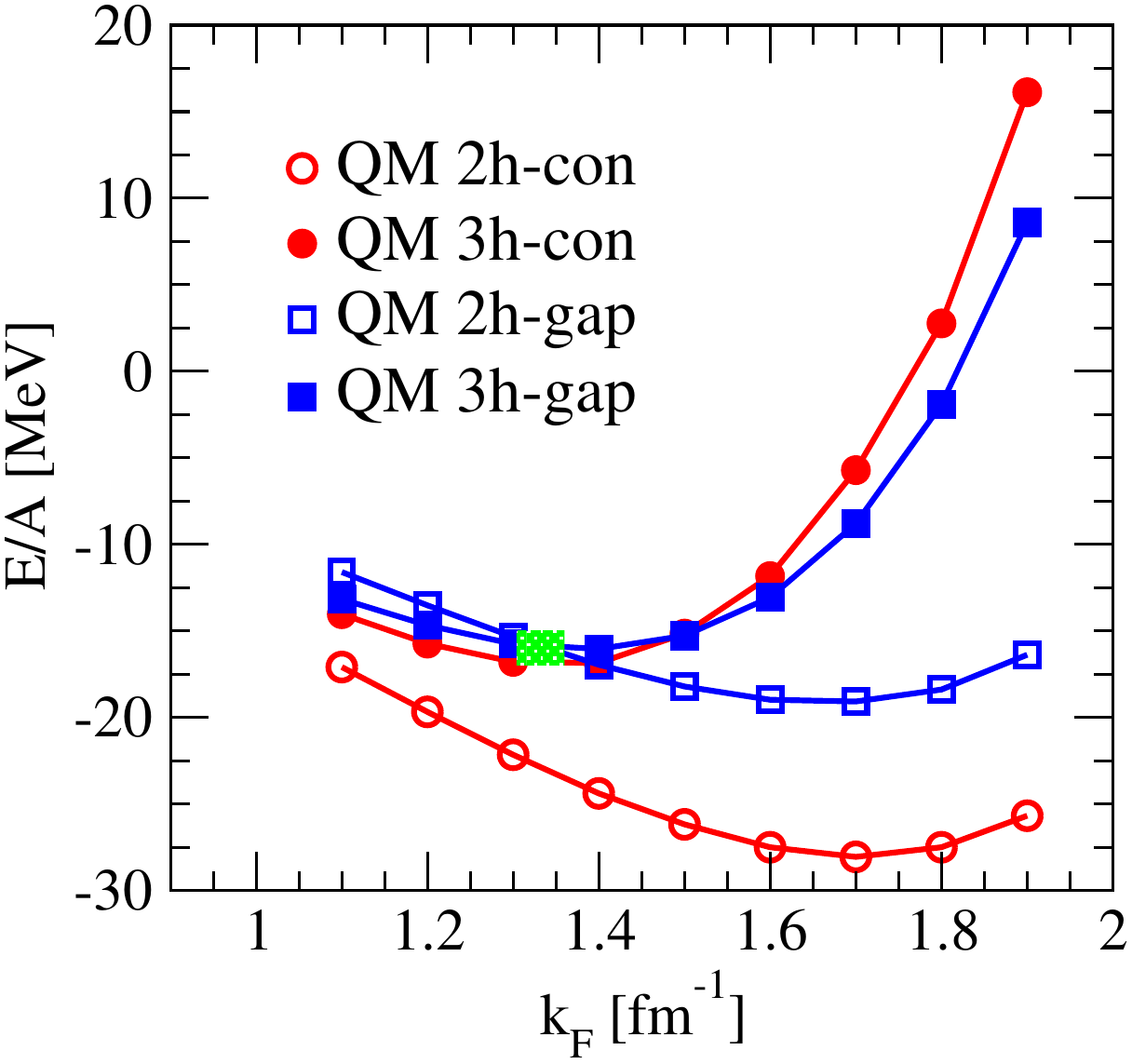}
\caption{Left panel: EoS of symmetric nuclear matter including third-order perturbative corrections based on chiral interaction at N$^3$LO level. The cut-off for TBFs is fixed at $\Lambda_{3N}$ = 2.0 fm$^{-1}$, while the cut-off $\Lambda$ of the (evolved) chiral two-body interaction is set equal to the two indicated values. Figure adapted from \cite{2011PhRvC..83c1301H}. 
Right panel: EoS of symmetric nuclear matter from the quark model (QM) interaction fss2. The open (full) symbols correspond to the two (three) hole-line calculations, respectively. The circles (squares) indicate the calculation with the continuous (gap) choice for the single-particle potential. In both panels, the green box indicates the saturation point.}
\label{fig:chir}
\end{figure}

The contribution of the TBF to the saturation mechanism is quite relevant when chiral forces are used. 
In fact, as illustrated in Fig.~\ref{fig:chir} (left panel), without TBF the EoS does not display an apparent saturation, and anyhow close to saturation density the TBF contribution is quite large, of several MeV, in contrast to the case of meson exchange interactions, where the TBF contribution around saturation is of the order of 1 MeV. 
These calculations are perturbative in character, as indicated in the labels, but this feature holds true also in more refined calculations. 
We notice the relevance of the momentum cut-off $\Lambda$, that is introduced in order to control the point interaction forces, that otherwise would produce a divergent contribution.   
In general the chiral two-body forces are evolved according to the RG method before they are employed in the many-body calculations, as in \cite{2011PhRvC..83c1301H}. 
The same procedure has been followed in the BHF calculations of \cite{2015PhRvC..91e4311S}, where a similar relevance of the TBF was found. 
The most sophisticated many-body calculation with chiral forces is probably the one of \cite{2014PhRvC..89a4319H}, where the coupled cluster method was employed up to a selected set of three-body clusters.  
In the latter paper it was also found that it is difficult with the same chiral forces to fit both the binding energy of few nucleon systems (H, $^3$He) and the saturation point. 
This feature is common to the meson exchange forces discussed above, for which the same difficulty was found. 
A similar conclusion was found in the BHF calculations of \cite{2016PhLB..758..449L}, where it is suggested to fit simultaneously the few-body binding energy and the saturation point. \par

More recently it has been shown \cite{Baldo:2014rda} that the fss2 interaction is able to reproduce also the correct nuclear matter saturation point without any additional parameter or need to introduce TBFs. 
This is illustrated in Fig.~\ref{fig:chir} (right panel), where the EoS for symmetric matter is reported. 
The open symbols correspond to the EoS calculated at the BHF level of approximation with the gap (GC, squares) and the continuous (CC, circles) choices, while the full symbols correspond to the EoS calculated by including the three hole-line contribution. 
One can see that also in this case the final EoS is insensitive to the choice of the single-particle potential. 
The main result of this calculation is that the saturation point is reproduced without the introduction of TBFs.
Note that this is the only two-body interaction that is able to reproduce with a fair accuracy both the binding energy of few nucleon systems and the saturation point of nuclear matter, without the need of TBFs. 

The conclusion one can draw from this rapid review of results with different forces is that the relevance of the TBFs is model dependent and that the explicit introduction of the quark degrees of freedom reduces strongly the relevance of the TBFs and allows to connect few-body systems to nuclear matter.  

In the following, we will only consider microscopic EoSs which fit correctly the saturation point. 
The considered set of EoS includes variational calculations (APR) \cite{apr}, BHF calculations with TBFs, both phenomenological \cite{Baldo:1997ag, Tara2013} and microscopically derived \cite{1989PhRvC..40.1040G, Zhou:2004br}, and relativistic Dirac-Brueckner calculations \cite{Fuchs}.
A comparison among these EoSs, along with some phenomenological EoSs, will be discussed in Sect.~\ref{sec:models-phenom}. 

\subsubsection{Phenomenological approaches}
\label{sec:models-phenom}

Phenomenological approaches make use of effective interactions instead of bare ones to treat dense matter, either homogeneous or clustered.
Most of these approaches rely on the (nuclear) EDF theory, that has proved to be successful in reproducing the properties of medium-mass and heavy nuclei \cite{bender2003, sr2007} but can also be applied to describe infinite systems, either inhomogeneous (like SN cores ot NS crusts) or homogeneous (like NS cores).
Indeed, nuclear EDFs presently provide a complete and accurate description of ground-state properties and collective excitations over the whole nuclear chart (e.g.~\cite{bender2003, lala2004}). 
Non-uniform (nucleonic) clustered matter, that is present at subsaturation density at relatively low temperatures, can be treated using various models, like the NSE model, liquid-drop type models, (semi-classical) Thomas-Fermi models, etc. 
On the other hand, a different approach to construct the (phenomenological) EoS is to use purely parameterized EoSs, that do not rely on any description of the NN interaction. 
An example is given by the piecewise polytropic EoS for nuclear matter of \cite{read2009}, while a metamodel for the nucleonic EoS inspired from a Taylor expansion around the saturation density of symmetric nuclear matter is proposed and parameterized in terms of the empirical parameters in \cite{margue2018a} and employed to analyse global properties of NSs in \cite{margue2018b}.

\paragraph{\bf Nuclear EDF / Mean-field approaches}
\label{sec:models-edf}
The density functional theory has been very successfully applied in various fields of physics and chemistry.
The advantage of this method is to recast the complex many-body problem of interacting particles (like nucleons) into an effective independent particle approach (see, e.g.,~\cite{bender2003, lacroix2010, duguet2014} for a review).
The total energy of the system is thus expressed as a functional of the nucleon number densities, the kinetic energy densities, and the spin-current densities, which are functions of the three spatial coordinates.
It has been proved that the exact ground state of the system can be obtained from an energy minimisation procedure (see~\cite{hohe1964, kohn1965} for the case of electron systems). 
The issue lies in the fact that the exact form of the functional itself is not known a priori. 
Therefore, one has to rely on phenomenological functionals, either relativistic, usually derived from a Langrangian, or non-relativistic, traditionally derived from effective forces of Skyrme or Gogny type.
In the nuclear context, this approach has been often referred to as the self-consistent (relativistic) mean-field Hartree-Fock method, or the Hartree-Fock+BCS and Hartree-Fock-Bogoliubov (HFB) methods if pairing is included (see, e.g.,~\cite{ringschuck, brinkbroglia}, and Chap.~8 in this book for details on pairing).
The EDFs depend on a certain number of parameters fitted to reproduce some properties of known nuclei and nuclear matter, as well as ab-initio calculations of infinite nuclear matter. 
The non-uniqueness of the fitting procedure and the choice of the experimental data used to fit the parameters have led to several different functionals, that may give very different predictions when applied outside the domain where they were fitted (see, e.g.,~\cite{gorcap2014}).
The situation is particularly critical for astrophysical applications, where extrapolations of nuclear masses are required for the description of the deepest regions of the NS crust, in SN cores, and in nucleosynthesis calculations.
However, the reliability of these EDFs for very neutron-rich systems can be partially tested by comparing their predictions for the properties of pure neutron matter with results obtained from microscopic ab-initio calculations.
Moreover, another question arises as whether the EDF parameters determined by fitting nuclear data at zero temperature can be reliably used when applying the EDFs at finite temperature.
Different studies have shown that the temperature dependence of the couplings is rather weak up to a few tens of MeV (e.g., \cite{moupan2009, fcpg2012proc, fedlen2015}), but it remains to be clarified whether these conclusions still hold at higher temperatures ($\gtrsim 100$~MeV) that can be reached in CCSNe or NS mergers (see Sect.~\ref{sec:finitet}).

\begin{itemize} 
\item \textit{Non-relativistic EDFs}. \\
Non-relativistic approaches usually start from an Hamiltonian $\hat{H}$ for the many-body system, $\hat{H} = \hat{T} + \hat{V}$, where $\hat{T} = \sum_i \hat{p}^2/2m_i$ is the kinetic term ($\hat{p}$ being the momentum operator and $m_i$ the mass of the species $i$) and $\hat{V}$ is the potential term.
The latter accounts for the two-body (pseudo)potential, that allows one to incorporate physical properties like effective masses.
Three-body interactions were included explicitely in the seminal work by Vautherin and Brink \cite{vb1972}, while most recent EDFs rather employ density-dependent terms that include in an effective way higher-order correlations.
However, these terms can generate some issues when implemented beyond mean field (e.g.,  \cite{bender2003}).
The total energy of the system $E$ can also be written in terms of only the EDF without knowing explicitly the underlying Hamiltonian, $E = \int d^3 r\ \mathcal{E_{\rm EDF}} + E_{\rm Coul}$, where $\mathcal{E_{\rm EDF}}$ is the energy functional that includes the kinetic energy density and the interaction term modelling the effective interaction among particles, and $E_{\rm Coul}$ is the Coulomb energy.
In calculations including pairing, the pair energy, $E_{\rm pair}$, has to be accounted for, and in finite nuclei the corrections for spurious motion, $E_{\rm corr}$, have to be subtracted (e.g., \cite{bender2003}).

The Skyrme-type effective interactions are zero-range density-dependent interactions and they are widely used in nuclear structure and in astrophysical applications since they allow for fast numerical computations.
Since the pioneer work of Skyrme~\cite{skyrme1956}, several extensions have been proposed (see, e.g.,~\cite{les2007, bender2009, chamel2009, margue2009, marsag2009, zal2010, fmdp2011, helle2012, margue2012, dav2015}), allowing to include and study, for example, the tensor part of the EDF, the spin-density-dependent terms, as well as a surface-peaked effective-mass term.
The accuracy in reproducing experimentally measured properties of finite nuclei has been greatly increased in recent well-calibrated Skyrme-type EDFs (see, e.g.,~\cite{gcp2010, gcp2013b, gcp2013a,  cham2015, gcp2016} for the most recent BSk models from the Brussels-Montreal collaboration, and \cite{washi2012, korte2014}).
Even though in some cases Skyrme forces may exhibit some instabilities and self-interaction errors (see, e.g., the discussions in \cite{cao2010, chamel2010b, cg2010, helle2013, navpol2013, pastore2014, pastore2015}), these can be cured with appropriate modifications of the EDF.
In Skyrme forces, usually the pairing interaction is specified separately, even if attempts to construct the pairing force starting from the same Skyrme interaction exist (e.g., \cite{doba1984}), although this results in more involving calculations.
Many Skyrme models have been recently compared against several nuclear matter constraints in \cite{dutra2012}.
However, most of the criteria chosen by the authors to discriminate among the different parameterizations are still matter of debate, particularly regarding the symmetry energy coefficients (see e.g.~\cite{epjasym}), and most of the constraints are known with large error bars (see also Sects.~\ref{sec:constraints-nuclear} and \ref{sec:esym}). 
Therefore, it might be premature to rule out some models on those basis (see e.g. \cite{stev2013}).

On the other hand, finite-range (density-dependent) interactions are generally derived from the Gogny interaction \cite{gogny1980}. 
For these EDFs, the same finite-range interaction has been generally employed for the pairing term.
However, this kind of EDFs are less widely used in astrophysics with respect to the Skyrme ones, because of the more involving numerical computations (see, e.g., \cite{gorhil2009, gorhil2016, hilgor2016}; see also \cite{selrio2014} for an analysis of different Gogny interactions and their predictions of the homogeneous-matter properties).

In addition to the Skyrme and Gogny effective interactions, other non-relativistic approaches have been developed.
The two-body separable monopole (SMO) interaction has been designed to be an effective interaction whose terms are separable in the space (and isospin) coordinates with parameters fitted to the properties of finite nuclei \cite{stev2001, stone2002}.
Other approaches include the three-range Yukawa (M3Y) type interactions \cite{nakada2003} and the local EDF developed, e.g., in~\cite{fay2000, fay2001} in which the self-consistent Gor'kov equations are solved to study nuclear ground-state properties.
More recently, new EDFs have been constructed within an approach inspired by the Kohn-Sham density functional theory \cite{baldo2010, bal2013}.
These Barcelona-Catania-Paris(-Madrid) (BCP and BCPM) EDFs have been derived by introducing in the functional results from microscopic nuclear and neutron-matter BHF calculations, and by adding appropriate surface, Coulomb, and spin-orbit contributions.
With a reduced number of parameters, these EDFs yield a very good description of properties of finite nuclei.

Nevertheless, a particular attention has to be paid when applying non-relativistic EDFs at high densities, where the EoSs based on these EDFs may become superluminal.

\item \textit{Relativistic mean-field (RMF) and relativistic Hartree-Fock (RHF) models}. \\
RMF models have been successfully employed in nuclear structure, to describe both nuclei close to the valley of stability and exotic nuclei (see, e.g., \cite{nik2011} for a review, and \cite{dutra2014, dutra2016e, dutra2016} for a recent comparison of different RMF parameterizations).
RMF models have been constructed based on the framework of quantum hadrodynamics (see, e.g.,  \cite{fetwal1971, wal1974, serot1992}).
The basic idea of these models is the same as for non-relativistic mean-field approaches: the many-body state is built up as an independent particle or quasiparticle state from the single-particle wave functions, which are, in this framework, four-component Dirac spinors.
A nucleus is thus described as a system of Dirac nucleons whose motion is governed by the Dirac equation. 
The NN interaction can be described as zero-range (point coupling), where the single-particle potentials entering the Dirac equations are functions of the various relativistic densities, or as finite-range interaction, in terms of an exchange of mesons through an effective Lagrangian $\mathcal{L}= \mathcal{L}_{\rm nuc} + \mathcal{L}_{\rm mes} + \mathcal{L}_{\rm int}$, where the different terms account for the nucleon, the free meson, and the interaction contribution, respectively. 
The isoscalar scalar $\sigma$ meson and the isoscalar vector $\omega$ meson mediate the long and short-range part of the interaction, respectively, in symmetric nuclear matter, while isovector mesons (like the isovector vector $\rho$ meson and the isovector scalar $\delta$ meson) need to be included as well to treat isospin-asymmetric matter.
It is also possible to reformulate the model in terms of the corresponding EDF.
The RMF total energy is then given by $E = \int d^3 r\ \mathcal{E_{\rm RMF}} + E_{\rm Coul}$, where $\mathcal{E_{\rm RMF}}$ includes the nucleon, meson, and interaction contributions.
As in non-relativistic EDFs, $E_{\rm pair}$ has to be included when accounting for pairing and the centre-of-mass correction has to be subtracted \cite{bender2003} in finite nuclei. 
The interaction term depends on the nucleon-meson coupling constants that are usually determined by fitting nuclei or nuclear-matter properties.
In particular, coupling to scalar mesons is needed to obtain a correct spin-orbit interaction in finite nuclei. 
However, in the RMF, spin-orbit splitting occurs without the recourse to an assumed spin-orbit interaction.
The Klein-Gordon equations for the meson fields, coupled to the Dirac equations for the nucleons, are solved self-consistently in the RMF approximation, where the meson-field operators are replaced by their expectation values in the nuclear ground state. 
However, for a quantitative description of nuclear matter and finite nuclei, one needs to include a medium dependence of the effective mean-field interactions accounting for higher-order many-body effects, analogously to non-relativistic EDFs. 
A medium dependence can either be introduced by including non-linear (NL) meson self-interaction terms in the Lagrangian, or by assuming an explicit density dependence (DD) for the meson-nucleon couplings.
The former approach has been employed in constructing several phenomenological RMF interactions, like the popular NL3~\cite{lala1997}, PK1, PK1R \cite{long2004}, and FSUGold~\cite{todd2005} (see, e.g., \cite{maslov2015} for a recent study with a NL Walecka model \cite{bogbod1977, serwal1986}). 
In the second approach, the functional form of the density dependence of the coupling can be derived by comparing results with microscopic Dirac-Brueckner calculations of symmetric and asymmetric nuclear matter or it can be fully phenomenological, with parameters adjusted to experimental data (see, e.g. the DD-RMF models of \cite{nik2002, long2004, typ2005, gog2008, rocavin2011, anttyp2015}).
The density dependence gives rise to the so-called rearrangement contributions which are essential for the thermodynamic consistency of the model.
Generalised (g)RMF models, which are an extension of the DD-RMF models, where the degrees of freedom of nucleons and (light) clusters are included in the Lagrangian, have been also formulated (see Sect.~\ref{sec:models-cluster}).

On the other hand, point-coupling models have been developed (see, e.g., \cite{niko1992, rusfur1997, burv2002b, nik2008, zhao2010}), recently reaching a level of accuracy comparable to that of standard meson-exchange effective interactions when applied for the description of finite nuclei.
Parameters of these models can also be constrained by $\chi$EFTs \cite{fin2003, fin2004, fin2006}.

However, these models do not explicitly take into account the antisymmetrisation of the many-body wave function.
Despite the computational more involving character of the finite-range interaction mediated by meson exchange, relativistic Hartree-Fock including exchange terms and relativistic HFB accounting for pairing have also been implemented (see, e.g., \cite{meng2006} for a review and the more recent \cite{long2007, long2010}).

A RMF model incorporating the internal quark structure of baryons is the \textit{quark-meson coupling} model.
This approach treats nucleons as bound states of three quarks and interacting via meson exchange.
In addition to standard mesons, pions are also included.
This model has been applied to study NS properties, e.g., in \cite{thom2013, whit2014}.

\end{itemize}

\begin{figure}[thb]
\includegraphics[scale=.38]{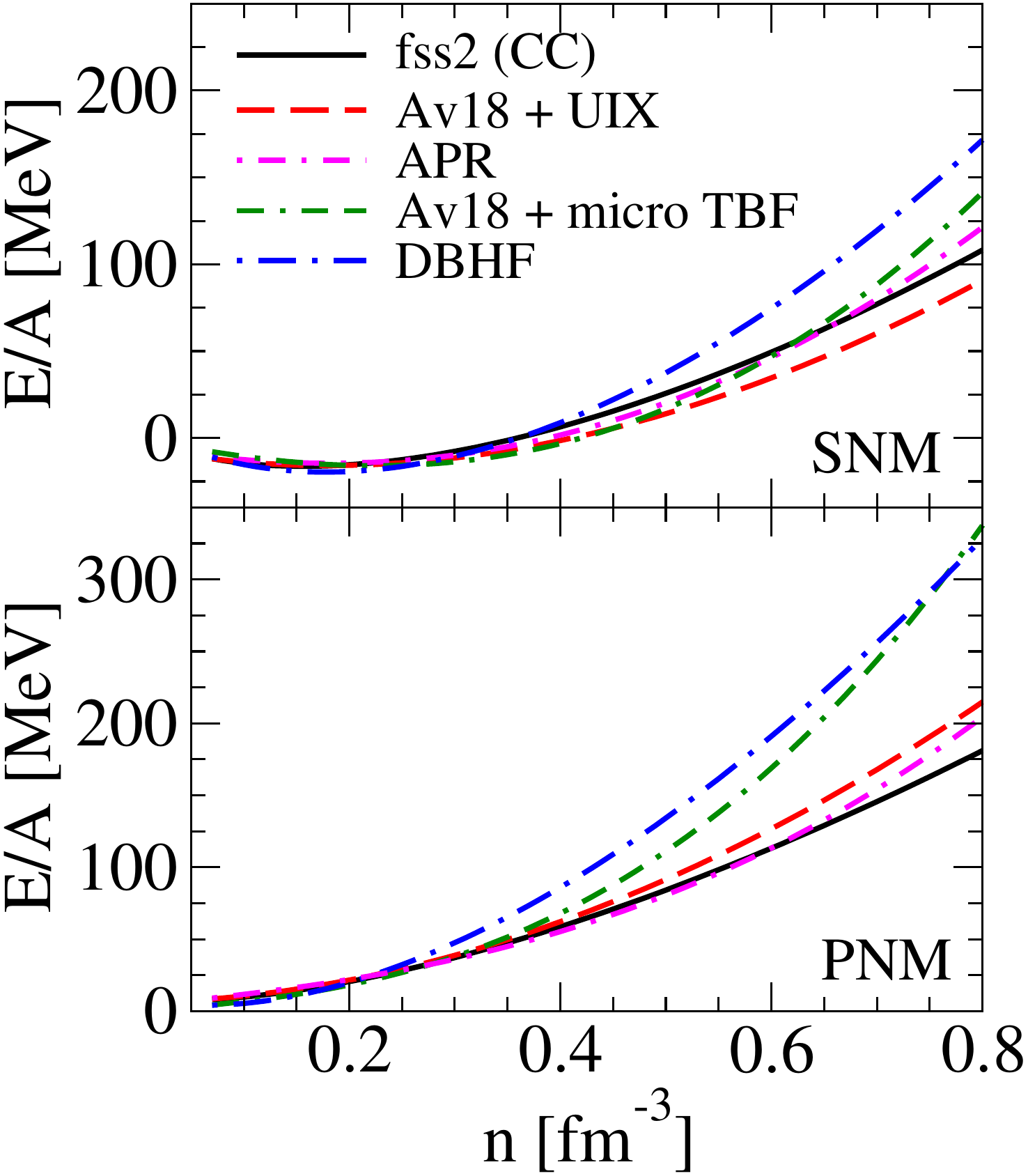}
\includegraphics[scale=.38]{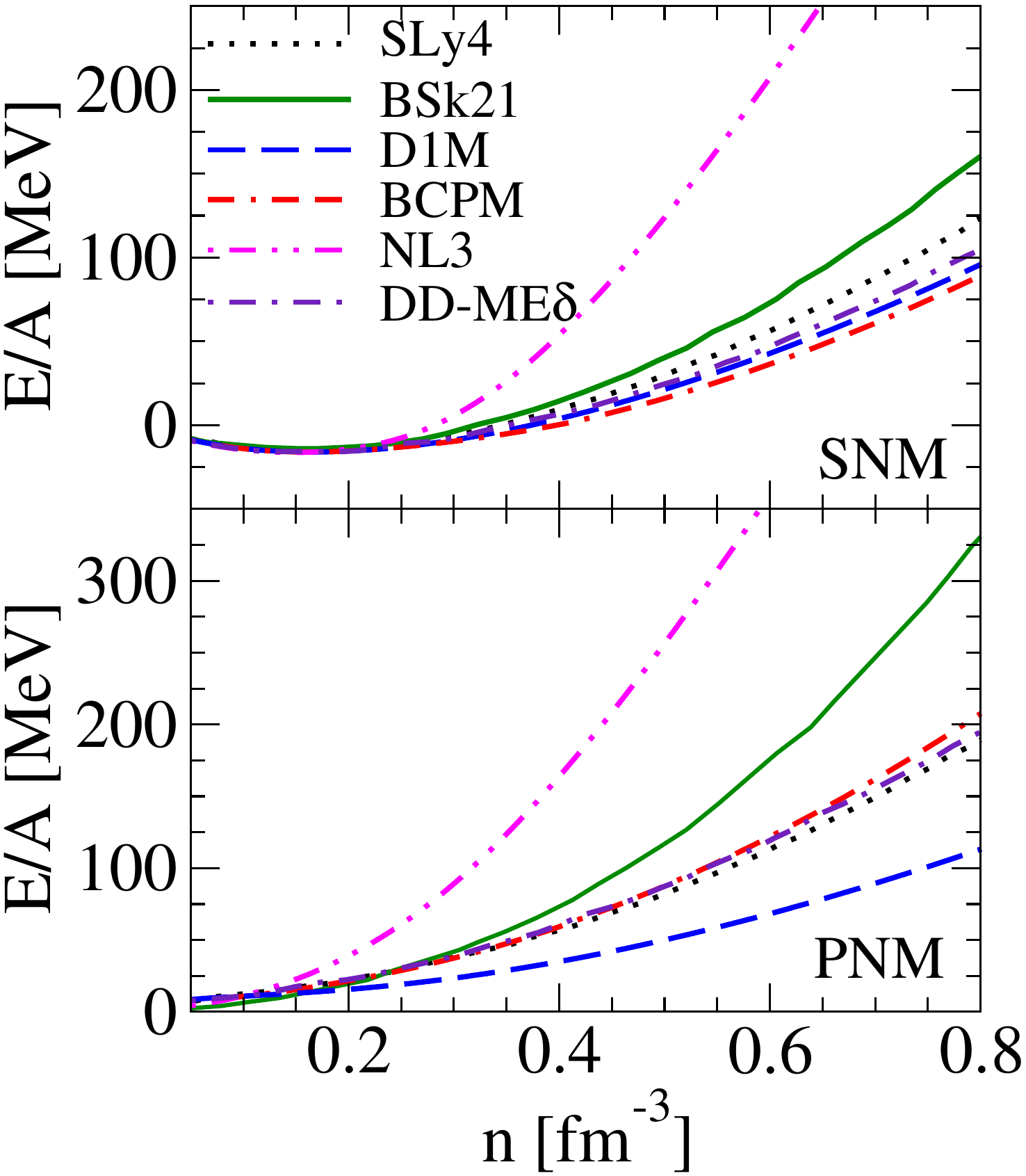}
\caption{Energy per particle in symmetric (upper panels) and pure neutron matter (lower panels) as a function of baryon number density for different models, both microscopic (left panels) and phenomenological (right panels). See text for details.}
\label{fig:pnm-snm}    
\end{figure}

As an illustrative example, in Fig.~\ref{fig:pnm-snm}, the energy per particle for symmetric (SNM, upper panels) and pure neutron matter (PNM, lower panels) is plotted as a function of baryon density, both for microscopic models (left panels) and for different phenomenological functionals (right panels).
Among the latters, we show the non-relativistic SLy4~\cite{chabanat1998} and BSk21~\cite{gcp2010} Skyrme-type EDFs, the D1M \cite{gorhil2009} Gogny EDF, and the BCPM~\cite{baldo2010, bal2013} functional, and the RMF NL3~\cite{lala1997} and DD-ME$\delta$~\cite{rocavin2011} models.
In both SNM and PNM, up to about twice the saturation density, all approaches, except very stiff EoSs, yield similar results.
Microscopic EoSs diverge at higher densities because of the different treatment of TBFs and three-body correlations.
Since microscopic calculations of PNM are very accurate, they can serve as benchmark calculations to constrain more phenomenological models.
The spread in these results can thus provide an estimate of the current theoretical uncertainties (see also \cite{ggc2015}).
For phenomenological approaches, a similar spread at high density can be noticed.
Indeed, these models have parameters that are fitted on experimental data known with some uncertainties around saturation (see also Sect.~\ref{sec:constraints-nuclear}), thus their behaviour at larger densities where no experimental data are available can be very different.
It has to be mentioned that, at low density, the appearance of clusters has to be considered in the EoS \cite{ropke2013}; the treatment of clustered matter will be discussed in the next Section.

\subsubsection{\bf Approaches to treat non-uniform matter}
\label{sec:models-cluster}

Non-uniform nuclear matter (either nuclei or clusters) is expected to be present at low densities (below saturation) and relatively low temperatures, thus in the crust of NSs and in SN cores.
At present, the best ab-initio many-body calculations employing realistic interactions are not affordable to describe inhomogeneous matter. 
Therefore, one has to rely on different approximations based on phenomenological effective interactions.
These approaches either (i) use the so-called single-nucleus approximation, i.e. the composition of matter is assumed to be made of one representative heavy nucleus (the one that is energetically favoured), possibly together with light nuclei (often represented by alpha particles) and unbound nucleons, or (ii) consider the distribution of an ensemble of nuclei.
It has been shown that employing the single-nucleus approximation instead of considering a full distribution of nuclei has a small impact on thermodynamic quantities \cite{burlat1984}.
However, differences might be significant if the composition is dominated by light nuclei, or in the treatment of nuclear processes like electron captures in CCSNe. 
Indeed, the nucleus that is energetically favoured from thermodynamic arguments might not be the one with the highest reaction rate. 
There are different ways to identify the onset of instability with respect to cluster formation, thus the transition from uniform to non-uniform matter (see, e.g., Landau and Liftshitz's textbook \cite{lanlif1980} and Chap.~7 of this book), although currently there exists no rigorous treatment to describe cluster formation beyond the single-nucleus approximation (see also Sect.~\ref{sec:finitet}).

As for electrons, in stellar environments like compact stars, they are usually treated as a non-interacting degenerate background gas (see, e.g., \cite{lattimer1996, hpy2007}).
In cold NS crusts, electron-charge screening (spatial polarisation) effects are small and the electron density is essentially uniform \cite{hpy2007, chamfant2016prd}; at densities $\rho_B \gg 10\ AZ$~g~cm$^{-3}$ ($\sim 10^4$~g~cm$^{-3}$ for iron, $A$ and $Z$ being the nucleus mass and proton number, respectively), the electrons can be treated as a quasi-ideal Fermi gas \cite{ch2008}.
For temperatures $T \gtrsim 1$~MeV and densities $\gtrsim 10^6$~g~cm$^{-3}$, leptons (electrons and neutrinos) are relativistic, in particle-antiparticle pair equilibrium and in thermal equilibrium with nuclear matter (see, e.g., \cite{lprl1985, ls1991}).

Nucleons can be either treated as a uniform system of interacting particles, or distributed within a defined shaped and sized cell.
In the latter case, often the Wigner-Seitz (WS) approximation is used: matter is divided in cells, each one charged neutral.
While at lower densities the cell is usually assumed spherical, centred around the positive charged ion surrounded by an essentially uniform electron and eventually free (unbound) nucleon (neutron and, at finite temperature, free proton) gas, at higher densities nuclei can be non-spherical and other geometries of the cell are considered. 
The standard way to calculate the EoS is then, for each thermodynamic condition, to minimise the (free) energy of the system with respect to the variational variables, e.g. the nucleus atomic and mass number, the volume (or radius) of the cell, and the free nucleon densities, under baryon number and charge conservation\footnote{Note that in the outer crust of cold catalysed NSs, the classical way to determine the EoS is to use the so-called BPS model \cite{bps1971}. In this model, the outer crust is supposed to be made of fully ionised atoms arranged in a body-centred cubic lattice at $T=0$ and to contain homogeneous crystalline structures made of one type of nuclides, coexisting with a degenerate electron gas (no free nucleons are present). The EoS in each layer of pressure $P$ is found by minimising the Gibbs free energy per nucleon, the only microscopic input being nuclear masses (see, e.g., \cite{hpy2007}).} (see, e.g., the pioneer work of \cite{bbp1971}). 
If additional structures, like the so-called ``pasta'' phases, are included, the minimisation is also performed on the shape of the cell (see, e.g., the pioneer works of \cite{rav1983, hashi1984}). 

Within the single-nucleus approximation, different models have been developed:
\begin{itemize}
\item \textit{(Compressible) Liquid-Drop Models}. \\
Liquid-drop models parameterize the energy of the system in terms of global properties such as volume, asymmetry, surface, and Coulomb energy; their parameters are fitted phenomenologically.
In these models, nucleons inside neutron-proton clusters and free neutrons outside are assumed to be uniformly distributed, and are treated separately. 
Moreover, clusters have a sharp surface, and quantum shell effects, despite playing a critical role in determining the equilibrium composition (particularly in the NS outer crust), are neglected. 
This approach has been among the earliest to be used in astrophysical applications to treat non-uniform matter at zero and finite temperature, because of its applicability and reduced computational cost (see, e.g., \cite{bbp1971, latt1981, lprl1985, ls1991, lorenz1993, wata2000, dh2000, dh2001, oya2007, naka2011}).

\item \textit{(Extended) Thomas-Fermi ((E)TF) models}. \\
These models allow for a consistent treatment of nucleons ``inside'' and ``outside'' clusters and are a computationally very fast approximation to the full Hartree-Fock equations.
The total energy of the system is written as a functional of the density of each species and their gradients.
Indeed, the (E)TF approximation to the energy density derived from a given nuclear EDF consists in expressing the kinetic-energy densities and the spin-current densities upon which the EDF depends as a function of the nucleon number densities (and their derivatives).
As a consequence, shell effects in the energy density are lost, but can be restored perturbatively using the Strutinski Integral (SI) theorem \cite{brack1985, onsi2008, pear2012, pear2015}. 
The density of nucleons in the cell can be either parameterized (e.g. using a Fermi-like profile) or obtained self-consistently.
These approaches have been developed in both non-relativistic and relativistic framework, at zero and finite temperature (see, e.g., \cite{sumi1995, cheng1997, shen1998, onsi2008, avan2010, shen2011, oka2012, pear2012, zhashe2014, sharma2015}).

\item \textit{Self-consistent mean-field models}. \\
Hartree-Fock models are fully quantum mechanical (see, e.g., \cite{nv1973, baldo2007, grill2011}). As a result, shell effects, which are found to disappear at temperatures above $2-3$~MeV, and pairing (in the Hartree-Fock+BCS and in the HFB approaches), which needs to be considered at temperatures below $1$~MeV~\cite{bq1974a}, are naturally included. 
However, these models are computationally very expensive and their current implementation is plagued by the occurence of spurious neutron shell effects \cite{chamel2007}.
Non-relativistic interactions are usually employed (e.g., \cite{nv1973, bv1982, mag2003, gogmut2007, newsto2009, paisto2012, papa2013, pais2014, sagert2016}), but RMF models have been also used \cite{maru2005}. 
\end{itemize}

At finite temperature, different configurations are expected to be realised.
A way to find these configurations is to solve the equations of motion and eventually exploit the ergodicity of the dynamics.
This can be done using time-dependent Hartree-Fock models and dynamical extended time-dependent Hartree-Fock approaches based on a wavelet representation (see, e.g., \cite{schue2013a} for the former models applied at finite temperature using Skyrme functionals - i.e. applicable to SN matter, and \cite{sebille2009, sebille2011} for the latter approach applied in the zero-temperature approximation - i.e. for NSs).
Another method is the \textit{classical molecular dynamics}, where nucleons are represented by point-like particles instead of single-particle wave functions.
So-called \textit{quantum molecular dynamics}, where nucleons are treated as wave packets, have also been developed.
However, in both cases particles move according to classical equations of motion and quantum effects (like shell effects) are not taken into account (see, e.g. \cite{wata2009, dorso2012, maru2012, lopez2014, schneider2014, caplan2015, gime2015, hor2015}, see also Chap.~7 in this book). 
However, these approaches are very time consuming.
On the other hand, one can assume that the system is in thermodynamic equilibrium and use NSE (or ``statistical'') models, where cluster degrees of freedom are introduced explicitly.
These models suppose that the system is composed of a statistical ensemble of nuclei and nucleons in thermal, mechanical, and chemical equilibrium.
The NSE is achieved when the characteristic time for nuclear processes is much shorter than the timescales associated to the hydrodynamic evolution of the system, and typically above $T \gtrsim 0.5$~MeV \cite{iliadis2007}.
Approaches considering an ensemble of nuclei are:

\begin{itemize}

\item \textit{(Extended) NSE}. \\
In the simplest version, NSE approaches treat the matter constituents as a mixture of non-interacting ideal gases governed by the Saha equation, where a Maxwell-Boltzmann statistics is employed, although quantum statistics (e.g. Fermi-Dirac for nucleons) can be incorporated.
The nuclear binding energies required as input of NSE calculations can be either experimental, whenever available \cite{audi2003, audi2012}, or theoretical (e.g. obtained from liquid-drop like models, or from more microscopic EDF-based mass models).
A limitation of standard NSE-based models is that they neglect interactions and in-medium effects, that are known to be very important in nuclear matter, especially at high densities.
For this reason, homogeneous matter expected to be present in NS cores, as well as the crust-core boundary in NSs, or matter at densities close to saturation density, cannot be correctly described by this kind of approaches, and microscopic or phenomenological models have to be applied instead.
Therefore, extended NSE models, where the distribution of clusters is obtained self-consistently under conditions of statistical equilibrium and interactions are taken into account, are developed.
For example, in-medium corrections of nuclear binding energies, either due to temperature or to the presence of unbound nucleons, have been calculated for Skyrme interactions in \cite{papa2013, aym2014} within a local-density and ETF approximation, respectively.
Some NSE models neglect the screening of the Coulomb interaction due to the electron background, while it is accounted for in other models, usually in the WS approximation (see, e.g., \cite{radgul2009, radgul2010, blinn2011}). 
The interactions between the cluster and the surrounding gas are often treated with an excluded-volume method (e.g., \cite{hs2010, radgul2010, hempel2012, furu2013, rag2014}).
However, the difference between the excluded-volume approach and the quantal picture proper of microscopic calculations leads to two different definitions of clusters in dense matter which in turn give differences in the observables \cite{papa2013}.
Excited states, that are populated at finite temperature, can be incorporated, either employing temperature-dependent degeneracy factors (as, e.g., in \cite{ios2003, radgul2010, gulrad2015}), or using temperature-dependent coefficients in the mass formula (as, e.g., in \cite{buy2014}).
Despite thermodynamic quantities are not very much affected by the presence of the ensemble of nuclei with respect to the single-nucleus approximation picture, quantitative differences arise in the matter composition, in particular concerning the contribution of light and intermediate mass nuclei (see, e.g., \cite{gulrad2015, rgo2016}).
Among the first applications of a NSE model for the EoS of SN cores at low densities is that of \cite{hnw1984,hw1985}.
NSE models have been subsequently employed for conditions encountered in CCSN, e.g., in \cite{botmis2010, hs2010, radgul2010, blinn2011, hempel2012, furu2013, buy2014, rag2014, gulrad2015, furu2017} (see also \cite{buy2013} for a comparison of methods).
As shown in Fig.~\ref{fig:NSE} for a typical condition encountered in the SN collapse, the NSE approach predicts a broad bi-modal distribution, centred around magic numbers.
This behaviour cannot be reproduced within the single-nucleus approximation, widely employed in SN simulations (see also Sect.~\ref{sec:eos-astro}).

Most of the aforementioned works make use of the excluded-volume approximation, which is less reliable for light nuclei, thus other approaches to treat interactions have been developed, as discussed below.

\begin{figure}[!ht]
\centering
\includegraphics[scale=0.4]{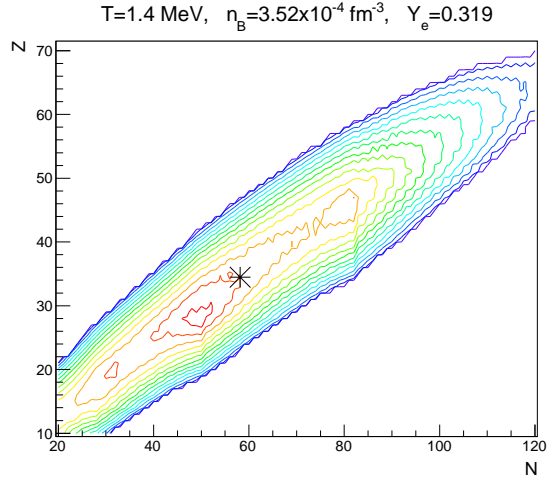}
\caption{Distribution of nuclei (blu to red: less to more abundant) for $n_B=3.52 \times 10^{-4}$~fm$^{-3}$, $T=1.4$~MeV, $Y_e=0.319$, calculated within the NSE model of \cite{gulrad2015}. The star corresponds to the result obtained in the single-nucleus approximation. Courtesy of Ad. R. Raduta.}
\label{fig:NSE}
\end{figure}

\item \textit{Virial EoS}.\\
The virial expansion, originally formulated by Beth and Uhlenbeck~\cite{uhlbeth1936, bethuhl1937}, is based on an expansion of the grand canonical potential in powers of the particle fugacities $z_i = \exp[(\mu_i-m_i c^2)/T]$, $\mu_i$ being the chemical potential of the particle $i$ and $m_i$ its mass.
It can thus be seen as an extension of NSE models to account for correlations between particles at low density and finite temperature.
This method relies on two assumptions: (i) the system is in a gas phase and has not undergone phase transition with decreasing temperature or increasing density, and (ii) the fugacity is small, so that the partition function can be expanded in powers of $z$.
The virial coefficients in the expansion are functions of temperature, and they are related to the two-, three-, and N-body correlations (see, e.g., \cite{bedaque2003, liu2009} for a discussion on coefficients beyond the second order).
In particular, the second virial coefficient is directly related to the two-body scattering phase shifts; thus, one can derive a model-independent approach up to $n_B \approx 10^{-5} - 10^{-4}$~fm$^{-3}$.
In the context of nuclear matter relevant for compact stars, the virial EoS has been applied, e.g., in \cite{horsch2006, horsch2006plb} to describe neutron matter and matter composed of nucleons and alpha particles, and in \cite{shenhor2010b, shenhor2011b} to model non-uniform matter at low densities.
However, this treatment is limited to light particles.
 
\item \textit{Models with in-medium mass shifts}. \\
In-medium mass shifts are a way to account for correlation effects in the medium, avoiding the use of an excluded volume. 
Nucleons and bound states (clusters) are treated on the same footing, as different constituent particles.
This approach also points out the appearance of the Mott effect due to Pauli blocking that prevents the formation of clusters at sufficiently high densities, and allows one to obtain the medium (density- and temperature-dependent) modification of cluster binding energies that enter into the EoS.
This in-medium modification approach has been included in different NSE-based models, like the \textit{quantum statistical model}, based on the thermodynamic Green's function method and developed, e.g., in \cite{ropke1982, ropke1983}; for recent applications of this approach to the description of light nuclei in nuclear matter at subsaturation densities, see, e.g., \cite{typ2010, ropke2011, ropke2013, ropke2015}. 
It has also been incorporated in the \textit{generalised (g)RMF models}, which are an extension of the DD-RMF models where nucleon and (light) cluster degrees of freedom are included in the Lagrangian (see, e.g. \cite{typ2010, vostyp2012, typ2013, typwol2014}). 
It turns out that the gRMF smoothly interpolates between the low-density virial EoS and the high-density limit of nucleonic matter, while the precise form of the transition depends, among other factors, on the choice of the coupling strength of the clusters to the meson fields.
A comparison of models using quantum statistical and gRMF models and the excluded-volume approach shows a rather good agreement at temperatures greater than a few MeV \cite{hemp2011}.
The effect of light clusters in RMF models in nuclear matter and in the pasta phase has been also investigated, e.g., in \cite{avan2012, pais2015}.

\end{itemize}

\subsection{Constraints on the equation of state}
\label{sec:constraints}

Theoretical models for the EoS can be constrained by both nuclear physics and astrophysical observations (see, e.g., \cite{pagereddy2006, klahn2006, latpra2007, trumper2011, tsang2012, lihan2013, latlim2013, latste2014b, stone2014, hor2014} for a discussion). 
However, in many cases, constraints on the EoS are not directly obtained from the raw data, but theoretical modelling is required to infer the constraints, or to extrapolate them in a region of the phase diagram not accessible by experiments or observations, thus making the constraints model dependent. 

\subsubsection{Constraints from nuclear physics experiments}
\label{sec:constraints-nuclear}

Valuable information on the many-body theories of nuclear matter is given by available data coming mainly from nuclear structure studies and heavy-ion collisions (HICs). 
Nuclear matter is an idealised infinite uniform system of nucleons, where the Coulomb interaction is switched off. 
Within the liquid-drop model of nuclei, if we put $E_{\rm Coul} = 0$ and in the limit of $A \rightarrow \infty$, the energy per nucleon, $E/A$, depends only on the neutron and proton densities, and because of charge symmetry of nuclear forces, it does not change if protons are replaced by neutrons and vice versa. 
Symmetric nuclear matter, with an equal number of neutrons and protons, is the simplest approximation to the bulk nuclear matter in heavy atomic nuclei. 
On the other hand, pure neutron matter is the simplest approximation to the matter as found in NS cores. 

As in the droplet model functional \cite{MS69}, it is convenient to express the binding energy in terms of the baryon density $n_B$ and the asymmetry parameter $\delta =  (N - Z)/A$, $N$ ($Z$) being the neutron (proton) number and $A=N+Z$. 
Usually, this energy is written as
\beq
E(n_B,\delta) = E(n_B,0) + S(n_B) \delta^2 \ ,
\label{eq:ebulk}
\eeq
$E(n_B,0)$ being the energy of symmetric nuclear matter ($\delta=0$) and $S(n_B)$ the symmetry energy. 
Both these terms can be expanded around the saturation density for symmetric matter, $n_0$, as 
\beqn
&& E(n_B,0) = E(n_0) \, +\, \frac{1}{18}\, K_0\, \epsilon^2 \ , \\
&& S(n_B) = S_0 \,+\, \frac{1}{3} L \epsilon \,+\, \frac{1}{18} K_{\rm sym} \epsilon^2 \ .
\eeqn
where $E(n_0)$ characterises the binding energy in symmetric nuclear matter at saturation,
$\epsilon = (n_B - n_0)/n_0$, $K_0$ is the incompressibility at the saturation point, $S(n_0) \equiv S_0$ is the symmetry energy coefficient at saturation, and the parameters $L$ and $K_{sym}$ characterise the density dependence of the symmetry energy around saturation. 
The value of $n_0$ and of the binding energy per nucleon for symmetric nuclear matter at saturation, $E(n_0)/A \equiv E_0/A$, can be extracted from experimentally measured nuclear masses, yielding $n_0 = 0.16 \pm 0.01$~fm$^{-3}$ and $E_0/A = -16.0 \pm 1.0$~MeV \cite{audi2012, 2013ADNDT..99...69A}. 
The uncertainties in these parameters result from the uncertainties in the experimental measurements and from the non-uniqueness of the fit of mass formulae used to reproduce thousands of nuclear masses. 
The coefficient $S_0$ determines the increase in the energy per nucleon due to a small asymmetry $\delta$, whereas the incompressibility $K_0$ gives the curvature of $E(n_B)$ at $n_B = n_0$ and the associated increase of the energy per nucleon of symmetric nuclear matter due to a small compression or rarefaction. 
These parameters are defined as :
\beq
K_0 \equiv 9 n_0^2 \left( \frac {\partial^2 E}{\partial n_B^2} \right)_{n_B=n_0, \delta=0} \,\,\ , \ \,\,
S_0 \equiv \frac{1}{2} \left( \frac {\partial^2 E}{\partial \delta^2} \right)_{n_B=n_0, \delta=0} \ ,
\eeq
and the higher-order symmetry energy coefficients, $L$ and $K_{\rm sym}$, are defined as
\beq
L \equiv 3 n_0 \left( \frac{\partial S(n_B)}{\partial n_B} \right)_{n_B=n_0} \,\,\ , \ \,\,
K_{\rm sym} \equiv 9 n_0^2 \left( \frac {\partial^2 S(n_B)}{\partial n_B^2} \right)_{n_B=n_0} \ .
\eeq

\begin{figure}[!h]
\centering
\includegraphics[scale=0.40]{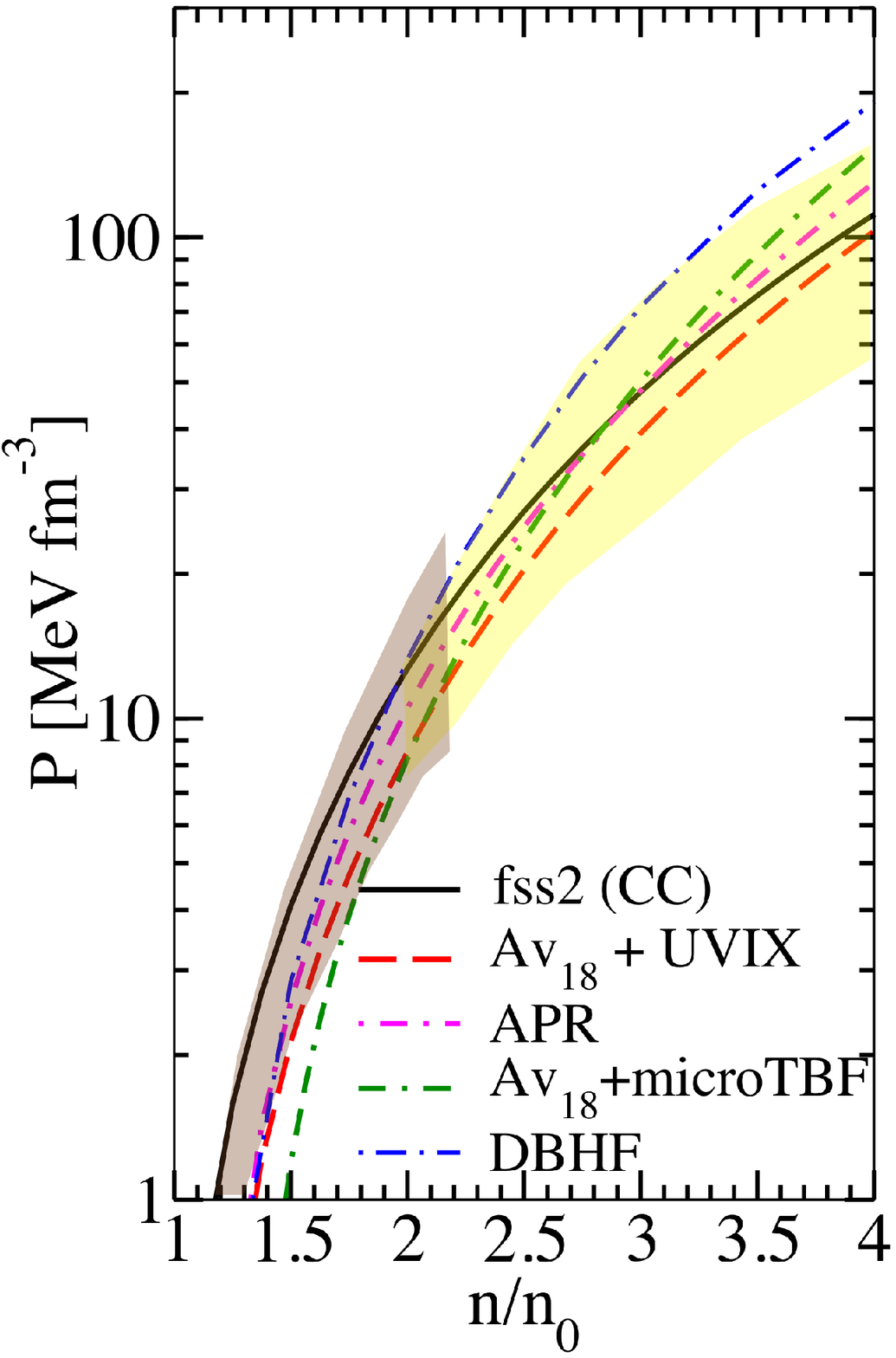}
\includegraphics[scale=0.40]{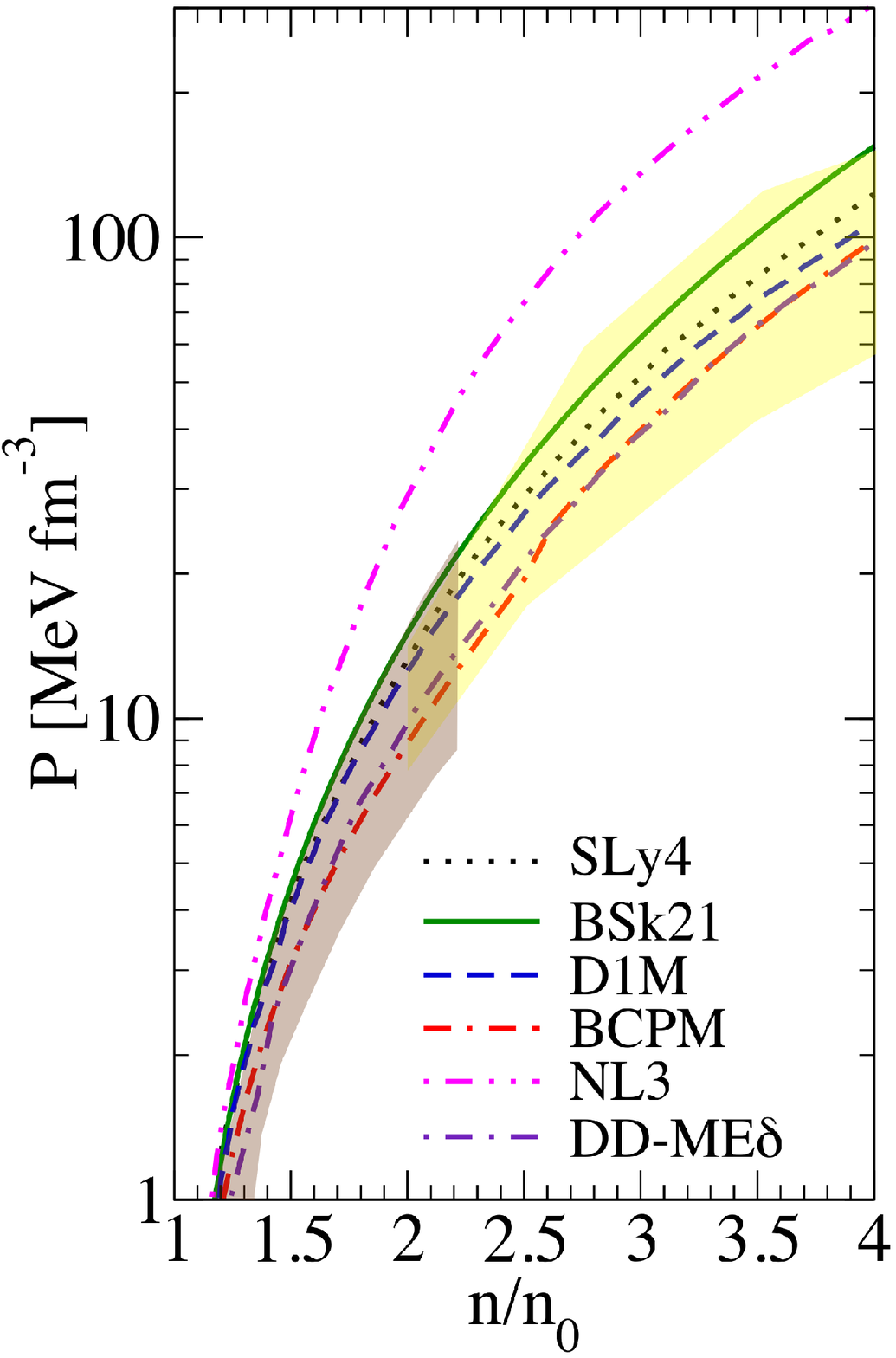}
\caption{Pressure versus baryon density (in units of saturation density) of symmetric nuclear matter for different microscopic (left panel) and phenomenological (right panel) models. 
The shaded area at lower (higher) density corresponds to constraints inferred from KaoS (flow) experiment \cite{daniel2002, lynch2009}.}
 \label{fig:Flow}
\end{figure}

The extraction of $K_0$ from experimental data is complicated and not unambiguous. 
Roughly speaking, RMF models predict larger values of $K_0$ with respect to non-relativistic EDFs (a large list of theoretical calculations of $K_0$ is given, e.g., in \cite{stone2014}).
Analysis of isoscalar giant monopole resonance in heavy nuclei suggests $K_0 = 240 \pm 10$~MeV \cite{colo2004} (a tighter constraint is reported in \cite{pieka2004}, $K_0 = 248 \pm 8$~MeV, while \cite{blaizot} gives $K_0 = 210 \pm 30$~MeV). 
However, it has been argued that data actually give information on the density dependence of the incompressibility around $0.1$~fm$^{-3}$ \cite{khanmar2013}.
HIC experiments (either flow experiments or kaon production experiments) would point to a rather ``soft'' EoS (e.g., \cite{fuchs2001, sturm2001, daniel2002, hartnack2006, lefevre2016}).
However, the inferred constraints remain model dependent since the data interpretation requires complex theoretical simulations (see also Sect.~\ref{sec:esym}). 
A discussion on the HIC constraints in relation with compact stars has been done, e.g., in \cite{sag2012, hempel2015}.
A comparison of the pressure versus density predicted by microscopic and phenomenological models with the results of the analysis on the flow and the kaon-production experiments \cite{daniel2002, lynch2009} is shown in Fig.~\ref{fig:Flow}.
Although not obvious, one can see that most of the EoSs, except very stiff ones, agree with these constraints (shaded area); only marginal deviations occur at the highest densities where the analysis is less reliable due to the possible presence of additional degrees of freedom or a phase transition. 
A refinement of the boundary could in principle put a more stringent constraint on the EoS, or even rule out some of them. 

\begin{figure}[!h]
\centering
\includegraphics[scale=0.40]{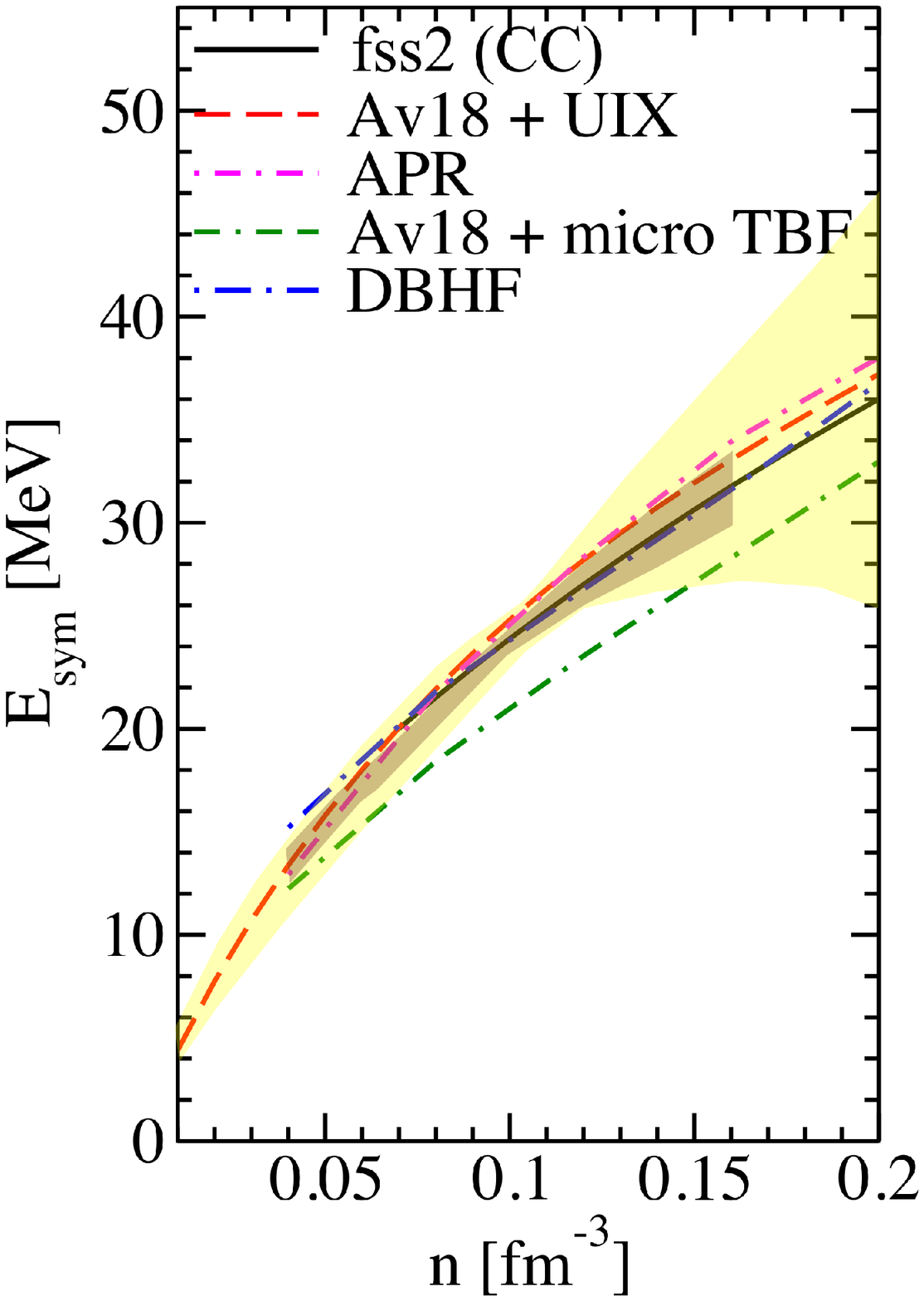}
\includegraphics[scale=0.40]{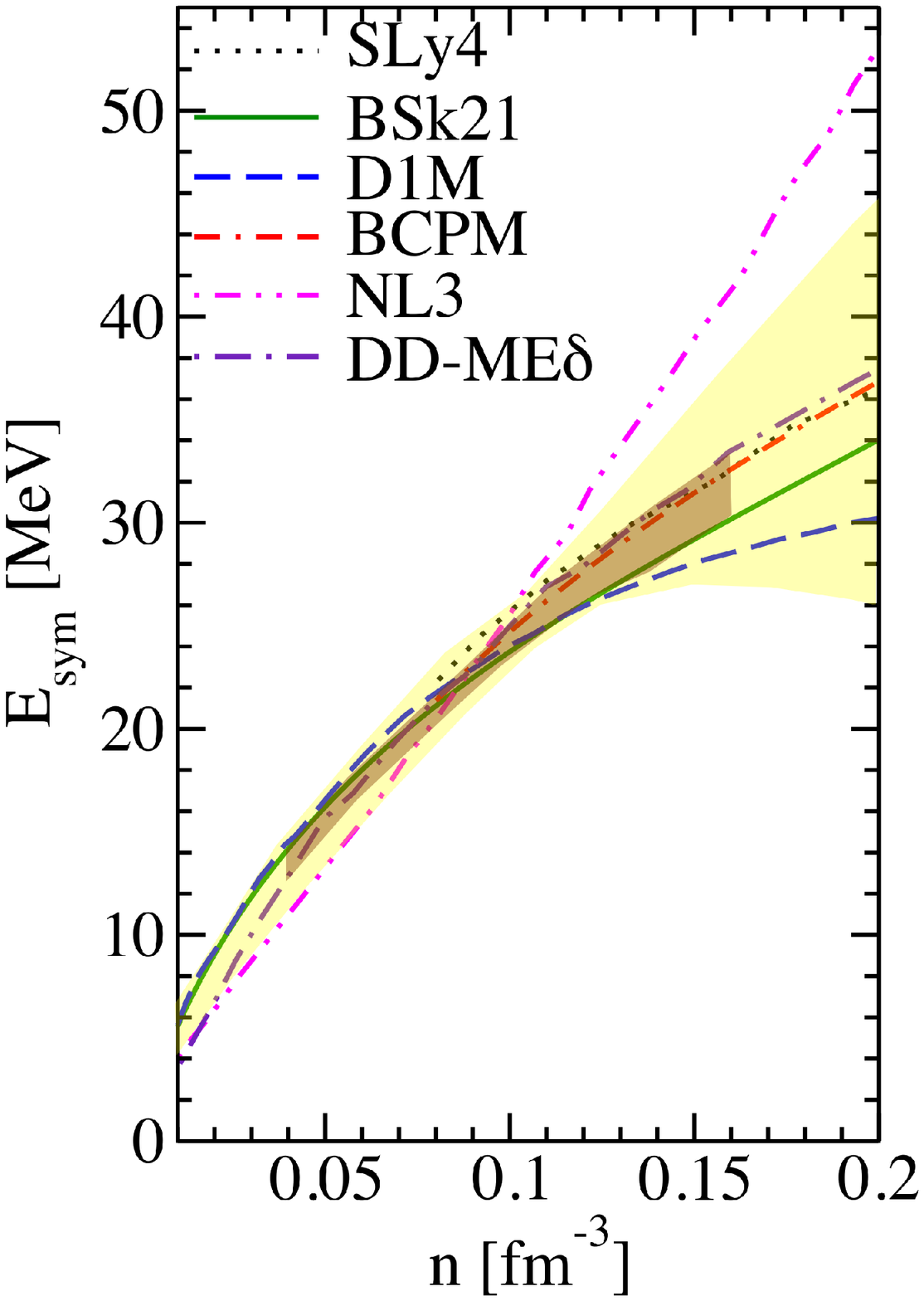}
\caption{Symmetry energy as a function of baryon number density calculated with different microscopic (left panel) and phenomenological (right panel) models.
The smaller (larger) shaded area corresponds to constraints inferred from the analysis of IAS (and corresponding extrapolation) of \cite{danlee2014}.} 
\label{fig:esym}
\end{figure}

Of particular importance for compact-star physics is the symmetry energy and its density dependence, which has been shown to affect the composition of NS crusts, the crust-core transition, and the neutron drip (see, e.g., ~\cite{epjasym, 2016PrPNP..91..203B} and references therein)\footnote{
It has also to be mentioned that not only the density dependence but also the temperature dependence of the symmetry energy, although not discussed here, can be important and may potentially have an impact in the CCSN dynamics (e.g., \cite{dpbb1994, dean2002, fbmmp2012, agra2014}).}.
The value of the symmetry energy at saturation, $S_0$, can be extracted, e.g., from nuclear masses, isobaric analog state (IAS) phenomenology, skin width data, and HICs; additional constraints come also from the NS data analysis.
Constraints on $L$ can be obtained, e.g., from the study of dipole resonances, electric dipole polarizability, and neutron skin thickness (see, e.g., \cite{lat2012, tsang2012, paar2014} and Sect.~\ref{sec:esym}).
While $S_0$ is fairly well constrained to lie around $30$~MeV, the values of the slope of the symmetry energy, $L$, and of higher order coefficients like $K_{\rm sym}$, at saturation, are still very uncertain and poorly constrained.
For example, combining different data, the authors of \cite{latlim2013} give $29.0 < S_0 < 32.7$~MeV, $40.5< L < 61.9$~MeV, while a more recent work suggests $30.2 < S_0 < 33.7$~MeV, $35 <L< 70$~MeV \cite{danlee2014}.
In Fig.~\ref{fig:esym}, we display the symmetry energy versus baryon number density for different microscopic (left panel) and phenomenological (right panel) models.
Note that for microscopic models, the curves of $E_{\rm sym}$ are given by the difference between the energy of pure neutron matter and that of symmetric matter, while for phenomenological models $E_{\rm sym}$ is calculated from the definition $E_{\rm sym}(n) = 1 / 2 (\partial^2 E / \partial \delta^2)|_{\delta=0}$.
Shaded areas represent constraints inferred from a study of the IAS and its extrapolation at lower and higher densities (see Fig.~15 in \cite{danlee2014}).
All the considered models, except very soft or very stiff ones, agree with these constraints. 
This means that the latters cannot be used to extract a simple functional parameterization of the density dependence of the symmetry energy. 
In particular, if one assumes a power law dependence, i.e. $ E_{\rm sym} \,\approx\, n^\alpha $, the exponential index $ \alpha $ cannot be constrained within a meaningful accuracy. 
Additional constraints, not reported here, have also been inferred at higher density, around and even beyond twice saturation density, from ASY-EOS and FOPI-LAND data (e.g., \cite{russotto2016}).

Finally, in Table~\ref{tab:nuc-par}, we list the nuclear parameters at saturation for the different microscopic and phenomenological models considered in the text, showing that all these models, except very stiff ones, agree reasonably well with the empirical values.

\begin{table}[!htb]
\centering
\caption{Nuclear parameters at saturation (saturation density $n_0$, energy per baryon $E_0/A$, incompressibility $K_0$, symmetry energy coefficient $S_0$, and slope of the symmetry energy $L$), for different microscopic and phenomenological models discussed in the text. Empirical values are taken from \cite{audi2012, 2013ADNDT..99...69A, colo2004, danlee2014}.}\smallskip
\label{tab:nuc-par}
\begin{tabular}{|c|ccccc|}
\hline 
EoS & $n_0$~[fm$^{-3}$] & $E_0/A$~[MeV] & $K_0$~[MeV] & $S_0$~[MeV] & $L$~[MeV] \\
\hline \hline
fss2 (CC)			& 0.157 & -16.3 & 219.0 & 31.8 & 52.0 \\
Av$_{18} + $ UVIX	& 0.16 & -15.98 & 212.4 & 31.9 & 52.9  \\
APR				& 0.16 & -16.0 & 247.3 & 33.9 & 53.8 \\
Av$_{18} + $ micro TBF	& 0.17 & -16.0 & 254.0 & 30.3 & 59.2 \\
DBHF			& 0.18 & -16.15 & 230.0 & 34.4 & 69.4  \\
\hline \hline 
SLy4				& 0.16 & -15.97 & 229.9 & 32.0 & 45.9 \\
BSk21 			& 0.158 & -16.05 & 245.8 & 30.0 & 46.6  \\
D1M				& 0.165 & -16.03 & 225.0 & 28.55 & 24.83 \\
BCPM 			& 0.16 & -16.0 & 213.75 & 31.92 & 52.96 \\
NL3				& 0.148 & -16.3 & 271.76 & 37.4 & 118.3 \\
DD-ME$\delta$ 	& 0.152 & -16.12 & 219.1 & 32.35 & 52.85 \\
\hline \hline
Empirical values	& $0.16 \pm 0.01$ & $-16.0 \pm 1.0 $ & $240 \pm 10$ & $30-34$ & $35-70$ \\
\hline
\end{tabular}
\end{table}
 
\subsubsection{Constraints from astrophysics}
\label{sec:constraints-astro}

Astrophysical observations can provide complementary information on the region of the dense-matter EoS which is not experimentally accessible in the laboratory.
These constraints mainly come from observations of NSs, either isolated or in binary systems (see also Chap.~5 of this book). 

\begin{itemize}
\item {\bf NS masses and radii}.
The most precise and stringent astrophysical constraints on the EoS come, at the present time, from the measurement of NS masses (see, e.g., \cite{ozelfreire2016} and the \texttt{nsmasses} website\footnote{https://stellarcollapse.org/nsmasses} for a recent compilation).
Indeed, the maximum mass of a NS is a direct consequence of general relativity and depends to a large extent on the high-density part of the EoS (above nuclear saturation density), where the EoS remains at present very uncertain. 
According to different calculations, the maximum mass of spherical non-rotating NSs is predicted to lie in the range $1.5 \msun \lesssim M_{\rm max} \lesssim 2.5 \msun$, $\msun$ being the mass of the Sun (see, e.g. \cite{cham2013, cham2013err} for a review). 
Recently, the mass of two NSs in binary systems have been precisely measured using the Shapiro delay: PSR J1614$-$2230~\cite{dem2010}, with a mass $M = 1.928 \pm 0.017\ \msun$ \cite{fon2016}, and PSR J0348$+$0432, with a mass $M = 2.01 \pm 0.04\ \msun$ \cite{anton2013}.
The latter mass is sufficiently high to put quite strong constraints on the EoS at densities four times larger than nuclear saturation, but it still remains compatible with a large class of models (see Fig.~\ref{fig:MR}).
However, this measured mass, which is about three times larger than the maximum mass of a star made of an ideal neutron Fermi gas, is a clear observational indication of the dominating role of strong interactions in NSs.
There also exist several less precise measurements of NS masses with values around and even above $ 2\ \msun$.
These measurements mainly refer to NSs in X-ray binaries or millisecond pulsar systems, where accretion, stellar wind, possible filling of Roche lobe by the companion, light-curve modelling, and other uncertainties could all play an important role. 
For this reason, the error of these mass measurements is quite large (see, e.g., \cite{hpy2007, iau2013} for a discussion). 

On the other hand, constraints on the EoS have been proposed using low-mass NSs.
Podsiadlowski et al. \cite{pod2005} suggested to probe the EoS using the observations of J0737$-$3039, a double pulsar system whose pulsar B has a mass $M = 1.2489 \pm 0.0007 \msun$~\cite{kramer2006}. 
The characteristics of the system suggest that pulsar B was formed after the accretion-induced collapse of an oxygen-neon-magnesium core that becomes unstable against electron capture.
Taking into account the uncertainties in the conditions of the pre-collapse core, it has been estimated that the critical baryonic mass of pulsar B for the onset of electron capture should be $1.366 < M_{\rm b} < 1.375 \msun$~\cite{pod2005}. 
If this scenario is correct, the knowledge of both the gravitational and the baryonic mass of pulsar B leads to a constraint on the EoS. 
However, there are various caveats in this analysis (e.g. neglect of mass loss during the SN, variation of the critical mass due to carbon flashes, formation history of this system), which can considerably change the constraint on the EoS \cite{pod2005, tauris2013}. 
In particular, Kitaura et al. \cite{kit2006} carried out hydrodynamical simulations of stellar collapse taking mass loss into account and found $M_{\rm b} = 1.360 \pm 0.002 \msun$.
A similar system has been observed recently, J1756-2251, where the mass of the pulsar is $1.230 \pm 0.007 \msun$ (one of the lowest NS mass measured with high accuracy)~\cite{ferd2014}.
However, the constraint inferred from the $M$ versus $M_{\rm b}$ relation strongly depends on the assumptions of the model and cannot therefore definitely rule out EoSs that do not satisfy it. 

A definite and stringent constraint on the EoS via the mass-radius relation would be the measurement of both mass and radius of the same object (see, e.g., \cite{read2009, stelat2013, ozefre2016b}).
However, precise estimations of NS radii are very difficult because more model dependent than those of masses,
mostly because observations of NS radii are indirect and the determination of the radius from observations is affected by large uncertainties (e.g., composition of the atmosphere, distance of the source, magnetic field, accretion; see, e.g., \cite{pot2014, fortin2015}).
Observations of the thermal emission from NSs can provide valuable constraints on their masses and radii.
The most reliable constraints are expected to be inferred from observations of transient low-mass X-ray binaries (LMXBs) in globular clusters because their distances can be accurately determined and their atmospheres, most presumably weakly-magnetised and primarily composed of hydrogen, can be reliably modelled.
Constraints can also come from observations of type I X-ray bursts, the manifestations of explosive thermonuclear fusion reactions triggered by the accretion of matter onto the NS surface.  
Recently, Steiner et al.~\cite{stelat2010, stelat2013} determined a probability distribution of masses and radii by analysing observations of type I X-ray bursters and transient LMXBs in globular clusters (see Fig.~\ref{fig:MR}).
However, this kind of analysis is still a matter of debate (see, e.g., \cite{lat2012, galdun2012, gui2013, guvoze2013, pout2014}).
Additional information on radii could also be inferred from X-ray pulsation in millisecond pulsars (see, e.g., \cite{bog2016}).

Future high-precision telescopes and missions like NICER, ATHENA$+$, and SKA are expected to improve our knowledge on the NS mass-radius relation (see, e.g., \cite{watts2015, bog2016b}).

In Fig.~\ref{fig:MR}, we display the gravitational mass $M$ versus radius $R$ for non-rotating NSs\footnote{It will be explained in Sect.~\ref{sec:eos-ns} how to construct a $M-R$ relation for a given EoS.}, obtained with different microscopic (left panel) and phenomenological (right panel) EoSs of asymmetric and beta-stable matter whose underlying models have been discussed in Sect.~\ref{sec:models}.
Note that only the EoSs based on the SLy4, BSk21, and BCPM EDFs are unified, while the others have been supplemented with an EoS for the crust (the BPS EoS \cite{bps1971}, except for the D1M model where the EoS of \cite{dh2001} was used).
Properties of NSs calculated with these EoSs are also reported in Table~\ref{tab:ns}.
These calculations assume that only nucleonic and leptonic (electrons and eventually muons) degrees of freedom are present inside the NS.
Horizontal bands correspond to the precise measurements of NS masses \cite{anton2013, fon2016}, while shaded areas correspond to the $M-R$ constraint inferred in \cite{stelat2013} (see their Fig.~1). 
Except very ``soft'' or ``stiff'' EoSs, the other considered EoSs are at least marginally compatible with the latter (model-dependent) constraint.
Instead, EoSs that predict a maximum NS mass below the observed ones have to be ruled out\footnote{Although rotation can increase the predicted maximum mass, this increase amounts to a few \% only even for the fastest spinning pulsar known (e.g., \cite{cham2013, fcpg2013}).}.
However, it would be premature to discard the underlying nuclear interaction as well.
Indeed, analyses of HIC experiments (e.g., \cite{fuchs2001, sturm2001, hartnack2006, xiao2009}) seem to favour soft (hadronic) EoSs.
This apparent discrepancy could be resolved by considering the occurrence of a transition to an ``exotic'' phase in NS cores (see, e.g., the discussion in \cite{cfpg2013}).
On the other hand, BHF microscopic calculations that include also hyperon degrees of freedom (e.g., ~\cite{2000PhRvC..61e5801B, 2011PhRvC..84c5801S, vidana2011}) show that the NS EoS becomes softer, and the value of the NS maximum mass is substantially reduced, well below the observational limit of $2\ \msun$. 
This poses a serious problem for the microscopic theory of NS interior (see the discussion in Chap.~7 of this book).

Another possibility to get a constraint on the mass-radius relation is to use observations of the gravitational redshift of photons emitted from the NS surface, $z_{\rm surf}$, a quantity related to the compactness ratio (proportional to $M/R$, see \cite{hpy2007}).
Spectroscopic study of the gamma-ray burst GRB 790305 from the soft gamma-ray repeater SGR 0526$-$066 \cite{higlin1990} suggested a value $z_{\rm surf} = 0.23 \pm 0.07$ for this object. 
Cottam et al. \cite{cot2002} also reported the detection of absorption lines in the spectra of several X-ray bursts from the LMXB EXO 0748$-$676, but this detection has not been confirmed by subsequent observations \cite{cot2008}.
Moreover, it has been argued that the widths of these absorption lines are incompatible with the measured rotational frequency of this NS \cite{lin2010}, but this point remains to be clarified \cite{baub2013}. 

Finally, the detection of neutrinos from SN1987A allows an estimation of the energy released during the SN core collapse.
Indeed, since 99\% of the energy of the SN was carried away by neutrinos of all flavours, the energy of the neutrino can be considered a measurement of the binding energy of the newly born NS. 
Defining the binding energy as the mass defect with respect to a cloud of iron dust \cite{dh2001} leads to a constraint on the NS gravitational mass. 
However, EoSs are usually found to be compatible with this constraint, thus making hard to rule out an EoS from it.

\begin{figure}[!th]
\includegraphics[scale=.33]{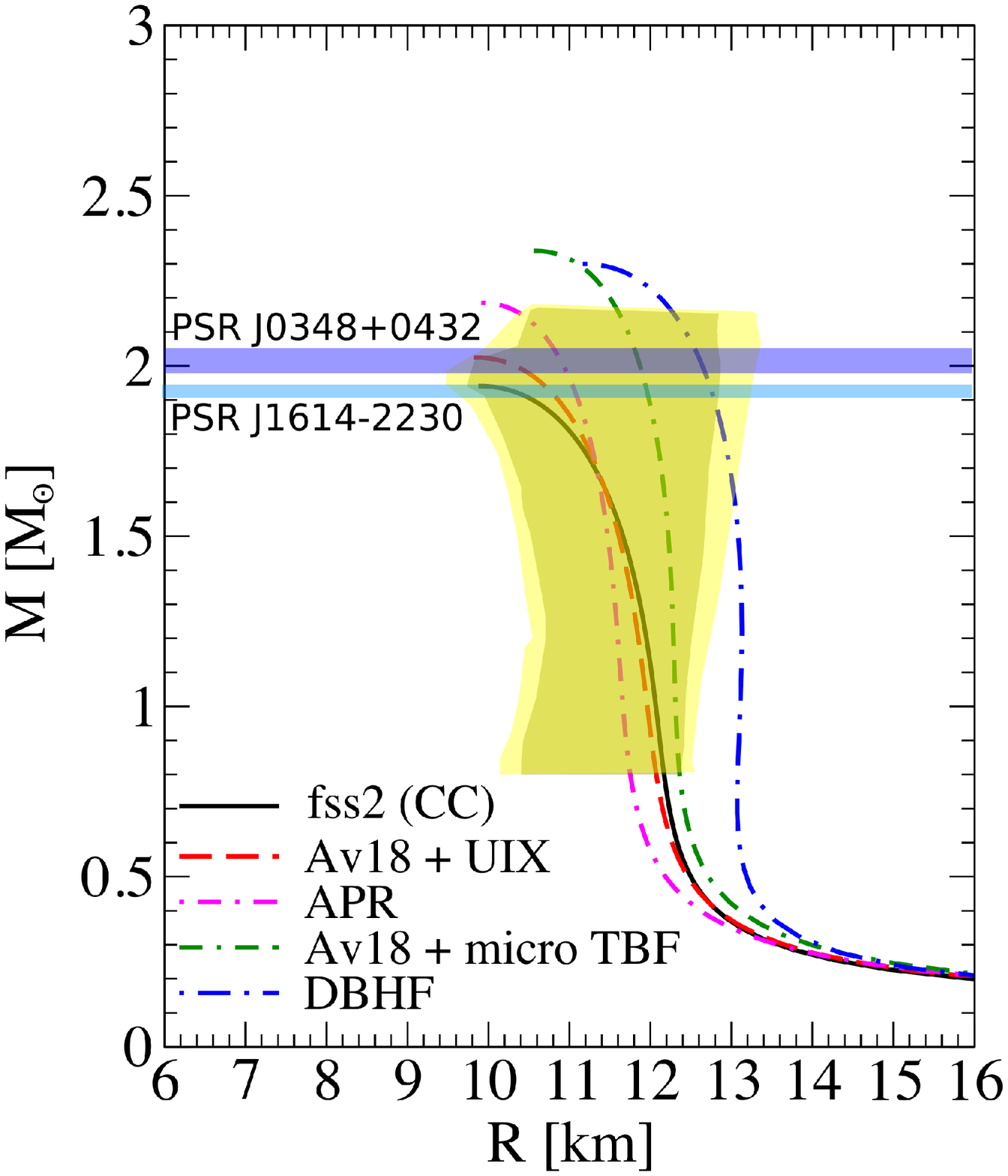}
\includegraphics[scale=.33]{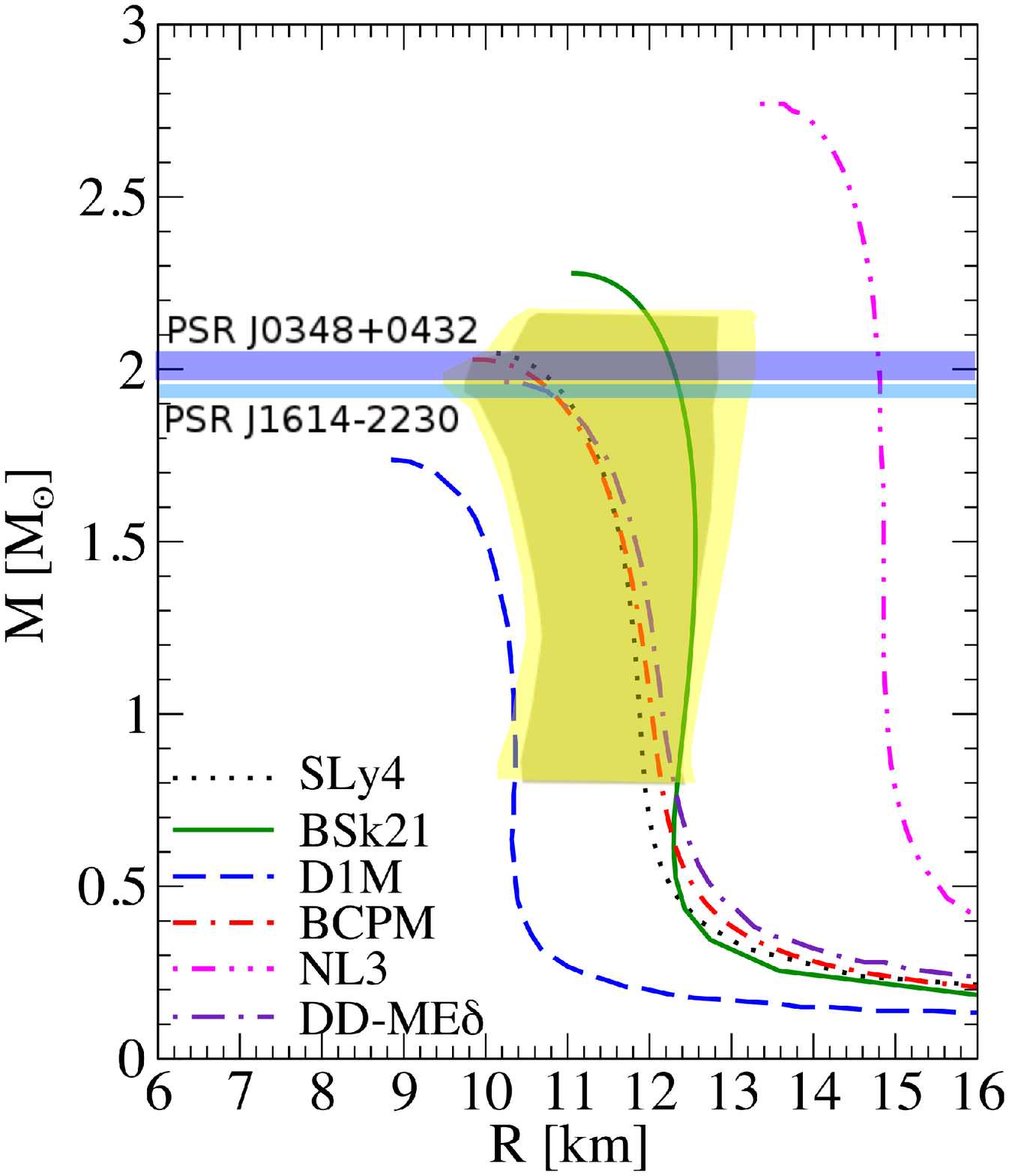}
\caption{Gravitational mass versus radius of non-rotating NSs for different nucleonic EoSs based on both microscopic (left panel) and phenomenological (right panel) models (results for the latters are taken from \cite{fcpg2013, gonzal2017, sharma2015, fatt2010, wang2014}). Horizontal bands correspond to the measured masses of PSR J$0348+0432$ \cite{anton2013} and PSR J$1614-2230$ \cite{fon2016}. Shaded areas correspond to the 68\% and 95\% confidences derived in \cite{stelat2013}.}
\label{fig:MR}    
\end{figure}

\begin{table}[!ht]
\centering
\caption{Properties of non-rotating NSs (maximum mass $M_{\rm max}$ and corresponding radius $R$, and central density $n_{\rm cen}$), for different microscopic and phenomenological models.}\smallskip
\label{tab:ns}
\begin{tabular}{|c|ccc|}
\hline 
EoS & $M_{\rm max}$~[$\msun$] & $R$~[km] & $n_{\rm cen}$~[fm$^{-3}$]  \\
\hline \hline
fss2 (CC)			& 1.94 & 9.9 & 1.87 \\   
Av$_{18} + $ UVIX	& 2.03 & 9.8 & 1.24   \\
APR				& 2.19 & 9.9 & 1.15  \\
Av$_{18} + $ micro TBF	& 2.34 & 10.6 & 1.01  \\
DBHF			& 2.30 & 11.2 & 0.97 \\
\hline \hline 
SLy4				& 2.05 & 10.0 & 1.21  \\  
BSk21			& 2.28 & 11.1 & 0.97  \\  
D1M 			& 1.74 & 8.9 & 1.57 \\     
BCPM			& 1.98 & 10.0 & 1.25 \\   
NL3				& 2.78 & 13.4 & 0.67  \\ 
DD-ME$\delta$	& 1.97 & 10.2 & 1.20 \\ 
\hline
\end{tabular}
\end{table}

\item {\bf NS rotation}.
Rotation of pulsars can be accurately measured.
The spin frequency of a NS must be lower than the Keplerian frequency, i.e. the frequency beyond which the star will be disrupted as a result of mass shedding.
Since the value of the Keplerian frequency obtained from numerical simulations of rotating NSs depends on the EoS (see, e.g., \cite{sterg2003} for a review), an observed frequency above the Keplerian one predicted for a given EoS would rule the EoS out.
Depending on the stiffness of the EoS, the highest possible rotational frequency for the maximum mass configuration has been found to lie in the range between $\sim 1.6$~kHz and $\sim 2$~kHz \cite{kra2008, haen2009, fcpg2013}.
Even the observation of PSR J1748$-$2446ad, the fastest spinning pulsar known \cite{hessels2006}, with a frequency of 716~Hz, cannot put stringent constraints on existing EoSs, because its rotational frequency still remains small compared to the Keplerian frequency.
Only observations of NSs with spin frequencies larger than about 1~kHz (see, e.g., \cite{bej2007, kra2008, haen2009}), could change the picture. 

\item {\bf NS cooling}.
Cooling observations are a promising way to probe the NS interior. 
Indeed, NS cooling depends on the composition and on the occurrence of superfluidity that determine heat transport properties (see, e.g., \cite{weber1999, yakpet2004, page2006, pot2015}, and Chaps.~8 and 9 in this book), thus potentially giving complementary information on the EoS.
For example, constraints on the mass-radius relation have been derived, using cooling models, from the observation of the central compact object in the SN remnant HESS J1731$-$347, that appears to be the hottest observed isolated cooling NS \cite{klo2015, ofe2015}. 
Recently, the impact of the stiffness of the EoS and in-medium effects on the cooling have also been studied \cite{gvb2016}.
A prominent role in the NS cooling is played by the neutrino emission due to the so-called direct URCA processes (e.g. \cite{tar2016}), which set in only if the proton fraction is larger than a certain threshold value. 
The proton fraction depends on the nuclear symmetry energy, and hence on the EoS. 
We will discuss this open question more in details in Sect.~\ref{sec:urca}.

\item {\bf NS moment of inertia}.
The mass and radius of a rotating NS can be constrained by measuring its moment of inertia $I$ (see e.g. \cite{latschu2005, wor2008, latpra2016}). 
Indeed, $I$ can be expressed as a function of the NS mass and radius (see e.g. the empirical formula for a slowly rotating NS proposed in \cite{latschu2005} and that holds for a wide class of EoSs, except for the very soft ones). 
Therefore, the radius could be determined if the mass and the moment of inertia of the NS is known.
However, the moment of inertia of a rotating NS has not yet been measured.
A lower bound can be inferred from the timing observations of the Crab pulsar, assuming that the loss of the pulsar spin energy goes mainly into accelerating the nebula (see, e.g., \cite{bejhaen2003, hpy2007}): only the EoSs predicting a value of $I$ higher than that estimated for Crab ($I \approx 1.4 - 3.1 \times 10^{45}$~g~cm$^2$ \cite{hpy2007, fcpg2013}) are acceptable.
However, the main uncertainty in this lower limit lies in the mass of the nebula, thus this constraint remains approximate.

\item {\bf Gravitational waves and oscillations}.
The merger of compact binary stars is expected to be the main source of the gravitational-wave signals observed with gravitational-wave detectors (see e.g. \cite{bairez2017} and also Chaps.~3, 10, 12 in this book).
It has been argued that the detection of gravitational waves from the post-merger phase of binary NSs could discriminate among a set of candidate EoSs (see e.g. \cite{baus2012, bsj2014, taka2014, taka2015, bsj2016, reztak2016}).
Also, quasi-periodic oscillations in soft gamma-ray repeaters could be used to derive constraints on the EoS (see e.g. \cite{stewat2009, sotani2012, gabler2013}).
This research area thus might be a promising way for constraining the EoS in the future (see e.g. \cite{bose2018, rezz2018} and Chap.~10 for a discussion on the effect of the EoS on the gravitational-wave signal from binary mergers)\footnote{During the refereeing process of this Chapter, the gravitational-wave signal from a binary NS merger, GW170817, has been observed in the galaxy  NGC 4993~\cite{abbott2017b}, in association with the detection of a gamma-ray burst (GRB 170817A) and electromagnetic counterparts~\cite{abbott2017b,abbott2017a}.}. 

\end{itemize}

\subsection{Applications to compact objects}
\label{sec:eos-astro}

In this Section we discuss some applications of the EoSs for compact objects, starting with the zero-temperature case relevant for NSs, then presenting some finite-temperature general purpose EoSs and their impact in compact-object simulations.

\subsubsection{Applications to neutron stars}
\label{sec:eos-ns}

\begin{figure}[ht]
\centering
\includegraphics[scale=0.43]{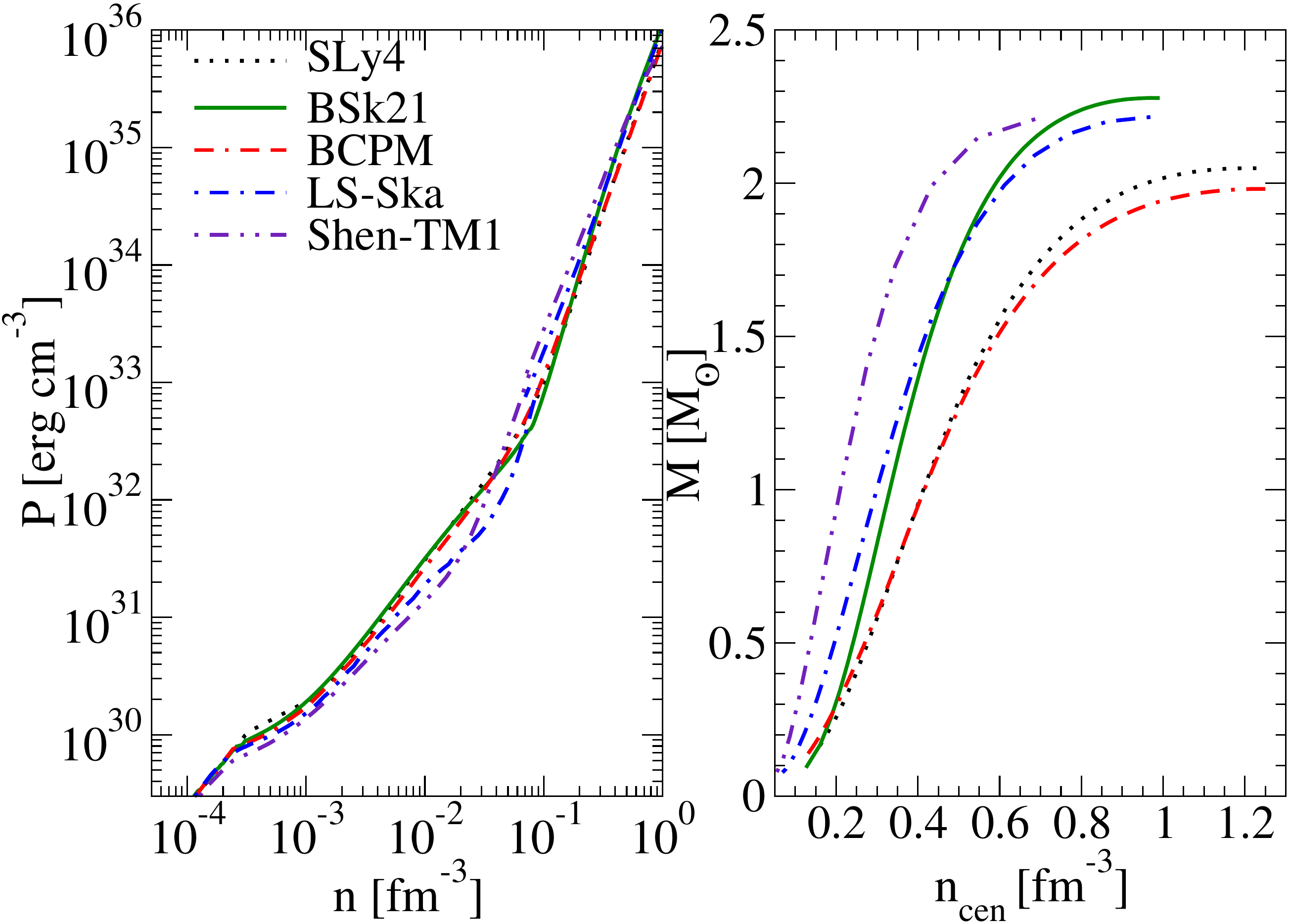}
\caption{Pressure versus baryon number density inside the NS (left panel) and NS gravitational mass versus central baryon density (right panel), for different unified EoSs.}
\label{fig:eos-unif}
\end{figure}

The NS physics and EoS have been extensively discussed, e.g., in \cite{hpy2007, ch2008, latpra2016} (see also Chap.~7 in this book).
Since the temperature in cold isolated NSs is below $\sim 1$~MeV, lower than characteristic nuclear Fermi energy, the zero-temperature approximation can be used in constructing the EoS.
The EoS is the necessary microphysics ingredient to determine the NS macroscopic properties, e.g. the mass-radius relation. 
Indeed, the structure of stationary, non-rotating, and unmagnetised NSs is determined by integrating the Tolman-Oppenheimer-Volkoff (TOV) equations for hydrostatic equilibrium in general relativity \cite{tol1939, ov1939} (see \cite{hpy2007} for details),
\begin{equation}
\frac{dP}{dr}=-\frac{G\rho \mathcal M}{r^2}\left(1+\frac{P}{\rho
c^2}\right)\left(1+ \frac{4\pi P r^3}{\mathcal M c^2}\right)\left(1-\frac{2G
\mathcal M}{r c^2}\right)^{-1}\, ,
\label{eq:tov1}
\end{equation}
where the function $\mathcal M(r)$ is defined by
\begin{equation}
\frac{d\mathcal M}{dr}=4\pi r^2 \rho\, ,
\label{eq:tov2}
\end{equation}
with the boundary condition $\mathcal M(0)=0$.
The gravitational mass of the NS is given by $M = \mathcal M(R)$, $R$ being the circumferential radius of the star where $P(R)=0$.
In order to solve these equations, an EoS, $P(\rho)$, must be specified.
The latter depends on the properties of dense matter which still remain very uncertain, especially in the core of NSs.
A number of EoSs for NSs are available, either with only nucleonic degrees of freedom, or with hyperonic and quark matter.
The majority of them are non-unified, i.e. they are built piecewise, matching different models, each one applied to a specific region of the NS.
On the other hand, in \textit{unified} EoSs, all the regions of the NS (outer crust, inner crust, and the core) are calculated using the same nuclear interaction (e.g., \cite{dh2001, fcpg2013, myn2013, sharma2015}). 
A unified and thermodynamically consistent treatment is important to properly locate the NS boundaries (such as the crust-core interface), that are important for the NS dynamics and which may leave imprints on astrophysical observables.
The use of non-unified EoSs may also lead to considerable uncertainties in the NS radius determination \cite{fortin2016}.
In Fig.~\ref{fig:eos-unif}, we show the NS pressure as a function of the baryon number density (left panel) and the NS mass versus the central density (right panel) resulting from the resolution of the TOV equations, Eqs.~(\ref{eq:tov1})-(\ref{eq:tov2}), for some unified EoSs. 
For those based on the SLy4 \cite{dh2001}, BSk21 \cite{fcpg2013}, and BCPM \cite{sharma2015} EDFs, the EoS of the outer crust is calculated in the standard BPS model \cite{bps1971};
for the former (SLy4), the outer-crust EoS is that of \cite{hp1994}, while for the latters nuclear masses are taken from the experimental data in \cite{audi2012} whenever available, complemented with theoretical mass models from HFB calculations with the corresponding functional.
For the inner crust, clusters are described within the compressible liquid-drop model in the EoS based on SLy4, thus no shell effects are included; however, different shapes of the WS cells (spheres, cylinders) are considered.
In the EoS based on the BSk21 EDF, the EoS for the inner crust has been calculated within the ETF model using a parameterized nucleon density distribution and with proton shell corrections included using the Strutinski Integral method, while in the BCPM EoS the self-consistent TF approach is employed and nuclear pasta is accounted for.
For the nucleonic liquid core, the EoSs are computed with the same functionals applied in the inner crust; note that for the BCPM EoS, the core EoS is derived in the framework of the BBG theory.
One can also obtain a unified EoS from a general purpose EoS, if applied at zero (or very small) temperature and in beta equilibrium.
For comparison, we display in Fig.~\ref{fig:eos-unif} the results for two interactions of such EoSs, that will be discussed in the next Section: the LS EoS in its SKa version\footnote{http://www.astro.sunysb.edu/lattimer/EOS/main.html} and the Shen EoS \cite{shen1998} with the TM1 parameter set.
All the considered EoSs predict a maximum mass around or above 2~$\msun$.
However, significant differences arise for the prediction of the central density and the radius of lower mass NSs.
Indeed, for a 1.5~$\msun$ NS, radii vary from $\sim 11.6$~km for the EoS based on the SLy4 EDF to $\sim 14.4$~km for the Shen-TM1 EoS (see, e.g., Fig.~16 in \cite{sharma2015}). 

\subsubsection{General purpose equations of state}
\label{sec:eos-genpurp}

Except for the case of ``cold'' (catalysed) NSs, for which the zero-temperature approximation can be used (see Sect.~\ref{sec:eos-ns}), for PNSs, CCSNe, and binary mergers, finite-temperature EoSs are crucially needed.
A detailed analysis of the finite-temperature properties of the bulk EoS relevant for CCSNe, PNSs, and binary mergers, has been done, e.g., in \cite{cos2014, cos2015}.
A wide range of densities, temperatures, and charge fractions, describing both clustered and homogeneous matter, is covered by the so-called ``general purpose'' EoSs.
These EoSs are therefore suitable for applications to SNe and mergers.
However, at present, only a few of them are available and direct applicable to simulations.
Moreover, for several of them, the underlying nuclear models are in disagreement with current constraints from either astrophysics (e.g. measured mass of NSs) or nuclear physics (experimental and/or theoretical constraints); see, e.g., the discussion in \cite{oertel2017} and Sect.~\ref{sec:constraints}.
We list below the general purpose EoSs with only nucleonic degrees of freedom currently used in astrophysical applications.
\begin{itemize}
\item \textbf{\textit{H\&W}}. The Hillebrandt and Wolff (H\&W) EoS \cite{hnw1984, hw1985} has been calculated using a NSE-network based on the model of \cite{eleid1980}, including $470$ nuclei in the density range $10^9 - 3 \times 10^{12}$~\gcm.
At higher densities, the EoS is computed in the single-nucleus approximation \cite{hnw1984}; the nuclear interaction employed is the Skyrme interaction with the SKa parameter set \cite{kohler1976}.
This EoS is still used in recent numerical simulations \cite{janka2012}.
\item \textbf{\textit{LS}}. The Lattimer and Swesty (LS) EoS \cite{ls1991} is a very widely used EoS in numerical simulations\footnote{The original EoS routine is available for three different parameterizations, according to the value of the incompressibility of the underlying nuclear interaction ($K_0=180$, $220$, and $375$~MeV), at http://www.astro.sunysb.edu/dswesty/lseos.html. More recent tables are given at http://www.astro.sunysb.edu/lattimer/EOS/main.html.}.
It models matter as a mixture of heavy nuclei (treated in the single-nucleus approximation), $\alpha$ particles, free neutrons and protons, immersed in a uniform gas of leptons and photons.
Nuclei are described within a medium-dependent liquid-drop model, and a simplified NN interaction of Skyrme type is employed for nucleons.
Alpha particles are described as hard spheres obeying an ideal Boltzmann gas statistics.
Interaction between heavy nuclei and the gas of $\alpha$ particles and nucleons are treated in an excluded-volume approach.
With increasing density, shape deformations of nuclei (non-spherical nuclei and bubble phases) are taken into account by modifying the Coulomb and surface energies, and the transition to uniform matter is described by a Maxwell construction.
\item \textbf{\textit{STOS}}. The Shen et al. (STOS) EoS \cite{shen1998, shen1998ptp, shen2011} is another widely used EoS.
As the LS EoS, matter is described as a mixture of heavy nuclei (treated in the single-nucleus approximation), $\alpha$ particles, and free neutrons and protons, immersed in a homogeneous lepton gas.
For nucleons, a RMF model with the TM1 interaction \cite{sugtoki1994} is used; $\alpha$ particles are described as an ideal Boltzmann gas with excluded-volume corrections.
The properties of the heavy nucleus are determined by WS-cell calculations within the TF approach employing parameterized density distributions of nucleons and $\alpha$ particles.
The translational energy and entropy contribution of heavy nuclei, as well as the presence of a bubble phase, are neglected (see \cite{zhashe2014} for a study of the effect of a possible bubble phase and a comparison between self-consistent TF calculations and those using a parameterized density distribution).
\item \textbf{\textit{FYSS}}. The EoS of Furusawa et al. \cite{furu2011, furu2013, furu2017} is based on a NSE model, including light and heavy nuclei up to $Z \sim 1000$.
For nuclei, the liquid-drop model is employed, including temperature-dependent bulk energies and shell effects \cite{furu2013, furu2017}.
For light nuclei, Pauli- and self-energy shifts \cite{typ2010} are incorporated \cite{furu2013, furu2017}.
The nuclear interaction used is the RMF parameterization TM1 \cite{sugtoki1994}.
The pasta phases for heavy nuclei are also taken into account.
The FYSS EoS has been applied in CCSN simulations to study the effect of light nuclei in \cite{furunaga2013}, and to investigate the dependence of weak-interaction rates on the nuclear composition during stellar core collapse in \cite{furu2017b}.
\item \textbf{\textit{HS}}. The Hempel and Schaffner-Bielich (HS) \cite{hs2010} EoS is based on the extended NSE model, taking into account an ensemble of nuclei (several thousands, including light ones) and interacting nucleons. 
Nuclei are described as classical Maxwell-Boltzmann particles, and nucleons are described within the RMF model employing different parameterizations.
Binding energies are taken from experimental data whenever available \cite{audi2003}, or from theoretical nuclear mass tables.
Coulomb energies and screening due to the electron gas are calculated in the WS approximation, while excited states of the nuclei are treated with an internal partition function, as in \cite{ios2003}.
Excluded-volume effects are implemented in a thermodynamic consistent way so that it is possible to describe the transition to uniform matter.
At present, EoS tables are available for the following parameterizations: TMA \cite{toki1995, hs2010, hempel2012}, TM1 \cite{sugtoki1994, hempel2012}, FSUgold \cite{todd2005, hempel2012}, NL3 \cite{lala1997, fischer2014}, DD2 \cite{typ2010, fischer2014}, and IU-FSU \cite{fatt2010, fischer2014}.
\item \textbf{\textit{SFHo, SFHx}}. The SFHo and the SFHx EoSs \cite{steiner2013} are based on the HS EoS, using two new RMF parameterizations fitted to some NS radius determinations.
These parameterizations have rather low values of the slope of the symmetry energy, $L$, with respect to those used in the HS EoS. 
\item \textbf{\textit{SHT, SHO}}. The EoSs of G.~Shen et al., SHT \cite{shenhor2011a} and SHO \cite{shenhor2011b}, are computed using different methods in different density-temperature domains\footnote{The SHT EoS is also available at the website http://cecelia.physics.indiana.edu/gang\textunderscore shen\textunderscore eos/.}. 
At high densities, uniform matter is described within a RMF model.
For non-uniform matter at intermediate densities, calculations are performed in the (spherical) WS approximation, incorporating nuclear shell effects \cite{shenhor2010a}.
The same RMF parameterization is employed.
In this regime, matter is modelled as a mixture of one average nucleus and nucleons, but no $\alpha$ particles \cite{shenhor2010a}.
At lower densities, a virial EoS for a non-ideal gas consisting of neutrons, protons, $\alpha$ particles, and $8980$ heavy nuclei ($A \geq 12$) from a finite-range droplet model mass table is employed \cite{shenhor2010b}.
Second-order virial corrections are included among nucleons and $\alpha$ particles, Coulomb screening is included for heavy nuclei, and no excluded-volume effects are considered.
In the SHT (SHO) model, the RMF NL3 (FSUGold) parameterization is used.
Since the original FSUGold EoS has a maximum NS mass of $1.7 \msun$, a modification in the pressure has been introduced at high density (above $0.2$~fm$^{-3}$), in order to increase the maximum NS mass to $2.1 \msun$. 
In order to produce a full table on a fine grid that is thermodynamically consistent, a smoothing and interpolation scheme is used \cite{shenhor2011a, shenhor2011b}.
\end{itemize}

Recently, EoSs at finite temperature incorporating additional degrees of freedom have been also developed.
Indeed, the appearance of additional particles such as hyperons, pions, or even a transition to quark matter, cannot be excluded in the density-temperature regime encountered during CCSNe or mergers.
\begin{itemize}
\item \textit{EoSs with hyperons and/or pions}. In the RMF framework, Ishizuka et al. \cite{ishi2008} extended the STOS EoS \cite{shen1998}, including hyperons and pions.
Scalar coupling constants of hyperons to nucleons are chosen to reproduce hyperonic potential extracted from hypernuclear data, while vector couplings are fixed based on symmetries.
These authors also investigate the impact of the EoS in NSs and in a spherical, adiabatic collapse of a $15 \msun$ star without neutrino transfer: hyperon effects are found to be small for the density and temperature encountered.
Pions are treated as an ideal free Bose gas.
Although, without interactions, charged pions condensate below some critical temperature, Ishizuka et al.~\cite{ishi2008} mention that pion condensation is suppressed when considering a $\pi N$ repulsive interaction.
Moreover, the effect of pions on the EoS is expected to become non-negligible at high temperature, where the free gas approximation should be valid. 
The EoS of \cite{ishi2008} has been applied, e.g., in \cite{naka2012} to investigate hyperons in BH-forming failed SNe.
An extension of the STOS EoS including pions is discussed, e.g., in \cite{naka2008}.
In the non-relativistic framework, Oertel et al.~\cite{ofn2012} added hyperons extending the model of Balberg and Gal \cite{bg1997}, that is based on a non-relativistic potential similar to that used in the LS EoS \cite{ls1991} for nucleons.
The hyperon couplings are chosen to be compatible with the single-particle hyperonic potentials in nuclear matter and with the measured NS mass of \cite{dem2010}.
Pions are also included in this model, as an ideal free Bose gas.
A version of this EoS, including only pions, has been employed to study BH formation \cite{peres2013}.

Other models, including only $\Lambda$ hyperons, have been also developed, e.g. extending the STOS EoS \cite{shen2011}, the LS EoS \cite{gul2013,peres2013}, or the HS model \cite{bhb2014} (an extended HS model with the DD2 interaction including hyperons and quarks with a constant speed of sound, $c_s^2=1/3$, has been considered in \cite{hein2016}).
The possibility of a phase transition at the onset of hyperons has been discussed, e.g., in \cite{schgal2000, schaf2002, gul2012, oertel2016}.
At low temperatures, the onset of hyperons occurs between about 2 and 3 times the saturation density. 
 The impact of additional particles on thermodynamic quantities (especially on the pressure) may be important for high temperature and densities (see, e.g., \cite{oertel2017}).
The role of hyperons in the dynamical collapse of a non-rotating massive star to a BH and in the formation and evolution of a PNS has been studied in \cite{banik2014} using the hyperonic STOS EoS~\cite{shen2011}.
\item \textit{EoSs with quarks}. Some EoSs also consider a phase transition to quark matter.
The MIT bag model is applied, e.g., in \cite{naka2008, naka2013, sag2009, sag2010, fischer2011, fischer2014b}, and the transition from hadronic to quark phase is modelled with a Gibbs construction.
The parameters of the model, the bag constant $B$ and the strange quark mass, impact the onset of the appearance of the quark matter.
The inclusion of a gas of pions raises the density of the transition to the quark phase due to the softening of the hadronic part of the EoS \cite{naka2008}.
The possible impact of a quark phase in the core-collapse dynamics will be briefly discussed below.
\end{itemize}

\paragraph{\bf Applications to core-collapse supernovae and black-hole formation}
\label{sec:eos-sn}

The main microphysics ingredients playing a crucial role in the CCSN dynamics are the EoS, the electro-weak processes (specifically, the electron capture on free protons and nuclei), and the neutrino transport (see also Chap.~1 in this book).
It has been shown that these inputs can have an important effect on the collapse dynamics and the shock propagation (see, e.g. \cite{bethe1990, mezza2005, janka2007, janka2012, burrows2013, janka2016} for a review).
However, their complex interplay and strong feedback with hydrodynamics make difficult to predict a priori whether a small modification of one of these inputs can have a considerable effect on the explosion (usually, effects are expected to be moderated, according to the Mazurek's law; see \cite{latpra2000, janka2012}).
In particular, the impact of the EoS is twofold: (i) it determines the thermodynamic quantities acting on the hydrodynamics (e.g. the pressure and entropy) and (ii) it determines the composition of matter thus affecting the electron-capture rates.
Concerning the latters, it has been shown that the single-nucleus approximation is not adequate to properly describe electron-capture rates during collapse.
Indeed, the most probable nucleus is not necessarily the one for which the rate is higher and this may have an impact on the $Y_e$ evolution thus on the collapse dynamics (see, e.g., \cite{lanmar2003, lanmar2014} for a review, and \cite{hix2003, sulli2016, rgo2016, rgo2017, furu2017b}).

Several studies have been carried out on the impact of the EoS on the infall and post-bounce phase, most in spherical symmetry, employing either EoSs in the single-nucleus approximation or based on a NSE approach (see, e.g. \cite{sumiyoshi2005, hempel2012, janka2012, steiner2013, fischer2014, toga2014b}).
Roughly speaking, a ``softer'' EoS would lead to a more compact and faster contracting PNS producing higher neutrino luminosities \cite{marek2009}, and to larger shock radii \cite{janka2012, suwa2013} in multi-dimensional simulations, resulting in a more favourable situation for explosion.
Different (``soft'' versus ``stiff'') EoSs may also potentially impact the gravitational-wave signal from SN (see, e.g. \cite{marek2009, schei2010, richers2017}).
However, it is not straightforward to correlate single nuclear parameters to the collapse dynamics, because different EoSs usually differ in many properties predicted by the underlying nuclear model and because spurious correlations between nuclear parameters can exist for a given model.
Moreover, other input parameters like the progenitor structure can impact the outcome of the simulations (see also Chap.~1 in this book).
Therefore, systematic investigations are difficult to perform, also because of computational costs of multi-dimensional simulations, and no strong conclusive statements can be drawn.

Since the EoS determines the maximum mass that the hot PNS can support, it also impacts the time from bounce until BH formation ($t_{\rm BH}$).
The sensitivity of $t_{\rm BH}$ on the EoS has been investigated, e.g., in \cite{sumi2007, sumi2007b,  fischer2009, oconn2011, ott2011}, using the LS and the STOS EoSs, and in \cite{naka2012, peres2013, banik2014, char2015}, where EoSs with additional degrees of freedom (hyperons, quarks, or pions) have been employed.
Especially in failed CCSNe, high temperatures and densities can be reached, so additional particles are expected to be more abundant.
It is generally found that the softening of the EoS thus induced reduces $t_{\rm BH}$, because the EoS supports less massive PNS with respect to the nucleonic EoS (see, e.g. \cite{naka2008, sumi2009, naka2012, peres2013, banik2014, char2015}). 

Some works have also claimed that a transition to a quark phase could have a non-negligible impact on the core-collapse dynamics (see, e.g., \cite{gentile1993, dratam1999, sag2009, fischer2011}).
In particular, it has been found that this phase transition can lead to a second shock wave triggering the explosion \cite{sag2009}.
Conditions for heavy-element nucleosynthesis in the explosion of massive stars triggered by a quark-hadron phase transition have also been investigated (e.g., \cite{fischer2011, nishi2012}).
However, the EoS applied by Sagert et al. \cite{sag2009}, based on the MIT bag model for the quark phase, was found to be in disagreement with the $2 \msun$ maximum mass constraint, and subsequent works could not systematically confirm the aforementioned scenario, leaving the question still open (see, e.g. \cite{sag2010, sag2012, naka2013, fischer2014b}).

\paragraph{\bf Applications to binary mergers}
\label{sec:eos-mergers}

Binary compact objects, either NSs or BHs, may also provide valuable information on the EoS of dense matter.
Indeed, they are promising sources of gravitational waves, they may produce short gamma-ray bursts (GRBs), and they are thought to be one of the main astrophysical scenarios for $r$-process nucleosynthesis (see, e.g., \cite{shitani2011, fabras2012, rosswog2015} for a review); all these scenarios are sensitive to the EoS\footnote{Several such studies have been conducted very recently, after the detection of the GW170817 event \cite{abbott2017b}.
The associated observations of the gamma-ray burst GRB 170817A and electromagnetic counterparts for this event suggest indeed that GW170817 was produced by the coalescence of two NSs followed by a short gamma-ray burst and a kilonova powered by the radioactive decay of $r$-process nuclei synthesised in the ejecta~\cite{abbott2017b, abbott2017a}.}.

Several studies show that the gravitational-wave frequency is related to the tidal deformability during the late inspiral phase of compact binary systems, and thus depends on the EoS (see, e.g., \cite{shitani2011, fabras2012, read2013, maselli2013, kumar2017}).
Moreover, the frequencies of the gravitational waves emitted during the post-merger phase are also sensitive to the NS EoS (see, e.g., \cite{seki2011, bsj2014, taka2014, bausster2015, palenzuela2015, taka2015, reztak2016, bairez2017}; see also Chap.~10 in this book).

It has also been proposed to probe the EoS using the analysis of short GRBs that are thought to be associated to binary-merger events (see, e.g., \cite{fanwu2013, lasky2014, fryer2015, law2015}).

Finally, the conditions and characteristics of $r$-process nucleosynthesis and the amount of ejected material depend of the thermodynamic conditions and matter composition of the ejecta thus on the EoS (see, e.g., \cite{goriely2011, bgj2013, wanajo2014}; see also Chap.~11 in this book).

\subsection{CompOSE and other online EoS databases}
\label{sec:compose}

CompOSE is an online database that has been developed within the European Science Foundation (ESF) funded ``CompStar'' network and the Europeean Cooperation in Science and Technology (COST) Action MP1304 ``NewCompStar''.
The database is hosted at the website http://compose.obspm.fr.
A manual describing how to use the database and how to include one's own EoS into it is also provided.
As stated on the main page of the website, ``The online service CompOSE provides data tables for different state of the art equations of state (EoS) ready for further usage in astrophysical applications, nuclear physics and beyond.''
It is not only a repository of EoS tables, but also provides a set of tools to manage the tables, such as interpolation schemes and data handling softwares.
At the time being, CompOSE hosts several one-parameter EoSs, suitable for application to NSs, and general purpose EoSs, applicable to SN matter.
More details and extensive explanations are given in \cite{typ2015}. 

Other online EoS databases that collect different available EoSs exist.
STELLARCOLLAPSE.ORG\footnote{http://www.stellarcollapse.org}, provides tabulated EoSs, as well as other resources for stellar collapse applications.
EOSBD\footnote{http://aspht1.ph.noda.tus.ac.jp/eos/index.html} aims ``to summarize and share the current information on nuclear EoS which is available today from theroretical / experimental / observational studies of nuclei and dense matter''.
The Ioffe website\footnote{http://www.ioffe.ru/astro/NSG/nseoslist.html} provides EoSs of fully ionised electron-ion plasma, EoSs and opacities for partially ionised hydrogen in strong magnetic fields, unified EoSs for NS crust and core, and some hyperonic EoSs; references to the original works are also given.
Relativistic EoS tables for SN are also provided online\footnote{http://user.numazu-ct.ac.jp/$\sim$sumi/eos/; \\ http://phys-merger.physik.unibas.ch/$\sim$hempel/eos.html}.

\section{Challenges and future prospects}
\label{sec:future}

\subsection{Model dependence of data extrapolations}
\label{sec:esym}

\begin{figure}[!h]
\centering
\includegraphics[scale=0.5]{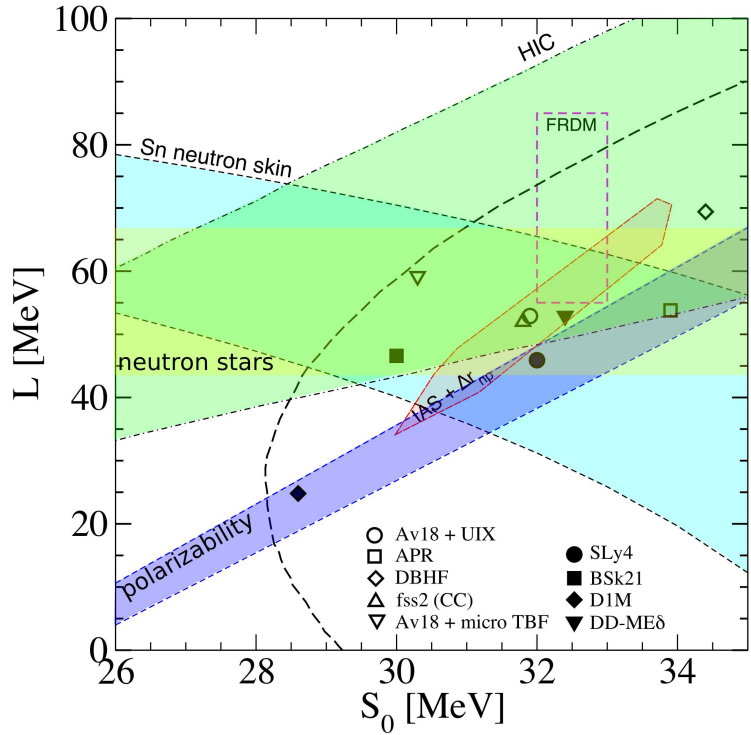}
\caption{Slope of the symmetry energy $L$ versus the symmetry energy coefficient $S_0$.
Shaded areas correspond to different experimental contraints. Symbols correspond to $S_0$ and $L$ predicted by different microscopic and phenomenological models.
See the text for details.}
\label{fig:L_S}
\end{figure}

One of the big issues of obtaining constraints on the EoS from experimental or observational data resides in the extrapolation of the raw data. 
Indeed, the majority of the constraints result from combining raw data with theoretical models, thus making the constraints model dependent. 
A typical example among astrophysical observations is the determination of NS radii (see Sect.~\ref{sec:constraints-astro} and Chap.~5 in this book).
Concerning constraints coming from nuclear physics experiments, issues arise since the state of matter in SNe and NSs is different compared to that in HICs: matter in SNe can be more isospin asymmetric and has to be charge neutral, while there is a net charge in HICs.
For example, the extraction of the pressure versus density constraint in symmetric nuclear matter shown in Fig.~\ref{fig:Flow} is subject to uncertainties of the transport models, which depend on a number of parameters that are not fully constrained.
Another important example is given by the inferred constraints on the symmetry energy. 
The latters are abundant at saturation density (see, e.g., \cite{tsang2012,latlim2013,latste2014b}). 
A (non complete) compilation of different experimental constraints is collected in Fig.~\ref{fig:L_S}, together with the values of $(S_0, L)$ predicted by different theoretical models, both microscopic (empty symbols) and phenomenological (filled symbols).
\par\noindent
i) The green shaded area marked as ``HIC'' corresponds to the constraints inferred from study of isospin diffusion in HICs \cite{tsang2009};
\par\noindent
ii) The turquoise shaded area labelled ``Sn neutron skin'' reports the constraints inferred from the analysis of neutron skin thickness in Sn isotopes \cite{chen2010};
\par\noindent
iii) The blue shaded area labelled ``polarizability'' represents the constraints on the electric dipole polarizability deduced in \cite{rocavin2015}.
In the latter work, available experimental data on the electric dipole polarizability, $\alpha_D$, of $^{68}$Ni, $^{120}$Sn, and $^{208}$Pb are compared with the predictions of random-phase approximation calculations, using a representative set of nuclear EDFs.
From the correlation between the neutron skin thickness of a neutron-rich nucleus and $L$, and between $\alpha_D S_0$ and the neutron skin thickness, Roca Maza et al. extracted a relation between $S_0$ and $L$ for the three nuclei under study (see Eqs.~(12)-(14) in \cite{rocavin2015}, and their Fig.~5).
The overlap of these constraints is shown in Fig.~\ref{fig:L_S};
\par\noindent
iv) The ``FRDM'' rectangle corresponds to the values of $S_0$ and $L$ inferred from finite-range droplet mass model calculations \cite{moller2012}.
These boundaries were derived by varying the considered sets of data along with different refinements of the model. 
Therefore, they can be biased by the uncertainties of the approach, and probably the constraints turn out to be too severe;
\par\noindent
v) The isobaric analog state (IAS) phenomenology and the skin width data can put tight constraints on the density dependence of the symmetry energy up to saturation.
These constraints give a range of possible values for $S_0$ and $L$, which are displayed in the ``IAS $+ \Delta r_{\rm np}$'' diagonal region, which represents simultaneous constraints of Skyrme-Hartree-Fock calculations of IAS and the $^{208}$Pb neutron-skin thickness \cite{danlee2014}.
\par\noindent
Finally, the horizontal band labelled ``neutron stars'' is obtained by considering the 68\% confidence values for $L$ obtained from a Bayesian analysis of mass and radius measurements of NSs~\cite{stelat2013}, while the dashed curve is the unitary gas bound on symmetry energy parameters of \cite{tews2017} (see their Eqs.~(24)-(25) with $Q_n=0$): values of $(S_0,L)$ to the right of the curve are permitted.
Constraints have also been derived from measurements of collective excitations, like giant dipole resonances (see, e.g., \cite{trippa2008, latlim2013, latste2014b}) and pigmy dipole resonances (see, e.g., the discussion in \cite{daougor2011, reinaz2013}).
However, we do not display the former constraint, whose band would largely superpose with the other constraints for $S_0 > 30$~MeV, and the latter, because of the large theoretical and experimental uncertainties.
Note that there is no area of the parameter space where all the considered constraints are simultaneously fulfilled.
This is likely to be due to the current uncertainties in the experimental measurements and to the model dependencies that plague the extraction of the constraints from the raw data. 
Although combining different constraints reduces the uncertainties in the $(S_0,L)$ parameter space, no definitive conclusion can be drawn and, except for models predicting a too high (or low) value of the symmetry energy parameters, no theoretical models can be ruled out a priori on this basis.
Finally, it has also to be clarified whether the derived correlations among different parameters and observables have a physical origin or are due to spurious correlations between the model parameters.

\subsection{Many-body treatment at finite temperature, cluster formation}
\label{sec:finitet}

A unified and consistent treatment of the different phases of matter, both at zero and finite temperature, is extremely challenging (see also Chap.~7 in this book). 
While either microscopic or phenomenological approaches are suitable to describe homogeneous matter, the correct description of cluster formation, and more generally of phase transitions, both at low and high densities and temperatures, is far from being a trivial task. 
Indeed, at present, there exist no consistent and rigorous treatment at zero and finite temperature of cluster formation beyond the single-nucleus approximation. 
In extended NSE models, interactions between a cluster and the surrounding gas are often treated in the excluded-volume approach, but from virial and quantal approaches it is found that cluster properties themselves are modified by the presence of a gas (e.g., \cite{horsch2006, typ2010, hemp2011}) and interactions among clusters should be also considered (e.g., \cite{typ2014}).
Moreover, (i) these in-medium effects are density and temperature dependent and (ii) with increasing temperature, excited states of nuclei become populated and need to be incorporated in the model.
The way of implementing them is not unique, and the different treatments lead to a considerable spread in the predictions of extended NSE models (see, e.g., \cite{buy2013}).

Another issue concerns the extension of the many-body methods and the extrapolation of their predictions, particularly at high density and temperature.
For example, the non-uniqueness of the fitting procedure of the EDF parameters and the choice of the experimental data used to fit the parameters have led to different EDFs, thus yielding a large spread in their predictions outside of the domain where the EDFs were fitted (e.g.,~\cite{gorcap2014}).
Especially for compact-object applications, this question can be critical, since extrapolations of nuclear masses are needed to describe the deepest regions of the NS crust and SN cores.
On the other hand, the nuclear interaction itself can be temperature dependent.
The temperature dependence of the EoS is very important for the physics of CCSNe, PNSs, and compact-star mergers, where densities larger than the saturation density and temperatures up to hundreds of MeV can be reached.
In particular, the stiffness of the EoS and the temperature dependence of the pressure can be crucial in determining the final fate of the CCSNe. 
Therefore, efforts should be devoted on the study of the EoS in the high-temperature regime. 
In Sect.~\ref{sec:models-abinitio}, the current state of the art of microscopic calculations of the finite-temperature EoS has been already discussed. 
The extension of those calculations at large temperatures is not a trivial task. 
For instance, in the Bloch-De Dominicis theoretical framework \cite{1959NucPh..10..509B}, on which the finite-temperature BHF approach is based, several higher order diagrams have to be included in the expansion, both at two and three hole-line level.
These additional contributions could have sizeable effects that are not straightforward to predict a priori. 
For phenomenological models, the question arises as whether the EDF parameters determined by fitting nuclear data at zero temperature can be reliably used when applying the EDFs at finite temperature.
Thermal properties of asymmetric nuclear matter have been investigated within a relativistic model, showing that the couplings are weakly dependent on temperature, up to a few tens of MeV \cite{fedlen2015}.
Similar conclusions can be deduced, e.g., from \cite{moupan2009, fcpg2012proc}. 
In the former work, where an extra term has been added to a Skyrme-type interaction, it has been shown that the temperature dependence of the couplings is weak up to about 30~MeV.
Also, a good agreement is obtained when comparing the free energy and pressure of nuclear matter for Brussels-Montreal Skyrme models with ab-initio calculations at finite temperature, up to 20~MeV (see Fig.~1 in \cite{fcpg2012proc}).
It remains to be determined whether these conclusions still hold at the highest temperatures ($\gtrsim 100$~MeV) that can be reached in CCSNe or binary mergers.

\subsection{Role of three-body forces}
\label{sec:tbf}

In Sect.~\ref{sec:models-abinitio} we have shown that a NN interaction based on quark degrees of freedom \cite{Baldo:2014rda,Ken} is able to reproduce at the same time the three-body properties, and the saturation point of nuclear matter without introducing TBFs, just using some parameters fitted on the NN phase shifts and deuteron properties.
These results were obtained by including in the many-body calculation the three-body correlations within the hole-line expansion of the BBG formalism,
indicating that the explicit introduction of the quark structure of the nucleons is relevant for the NN interaction. 
Additional interactions based on quark degrees of freedom should be considered, in order to understand if they have similar properties and eventually to pinpoint the key reasons of their performance, which is comparable with that of the best NN interaction based on meson exchange processes or on the chiral symmetry of QCD.

\subsection{Composition and URCA process}
\label{sec:urca}

\begin{figure}[!htb]
\centering
\includegraphics[scale=0.35]{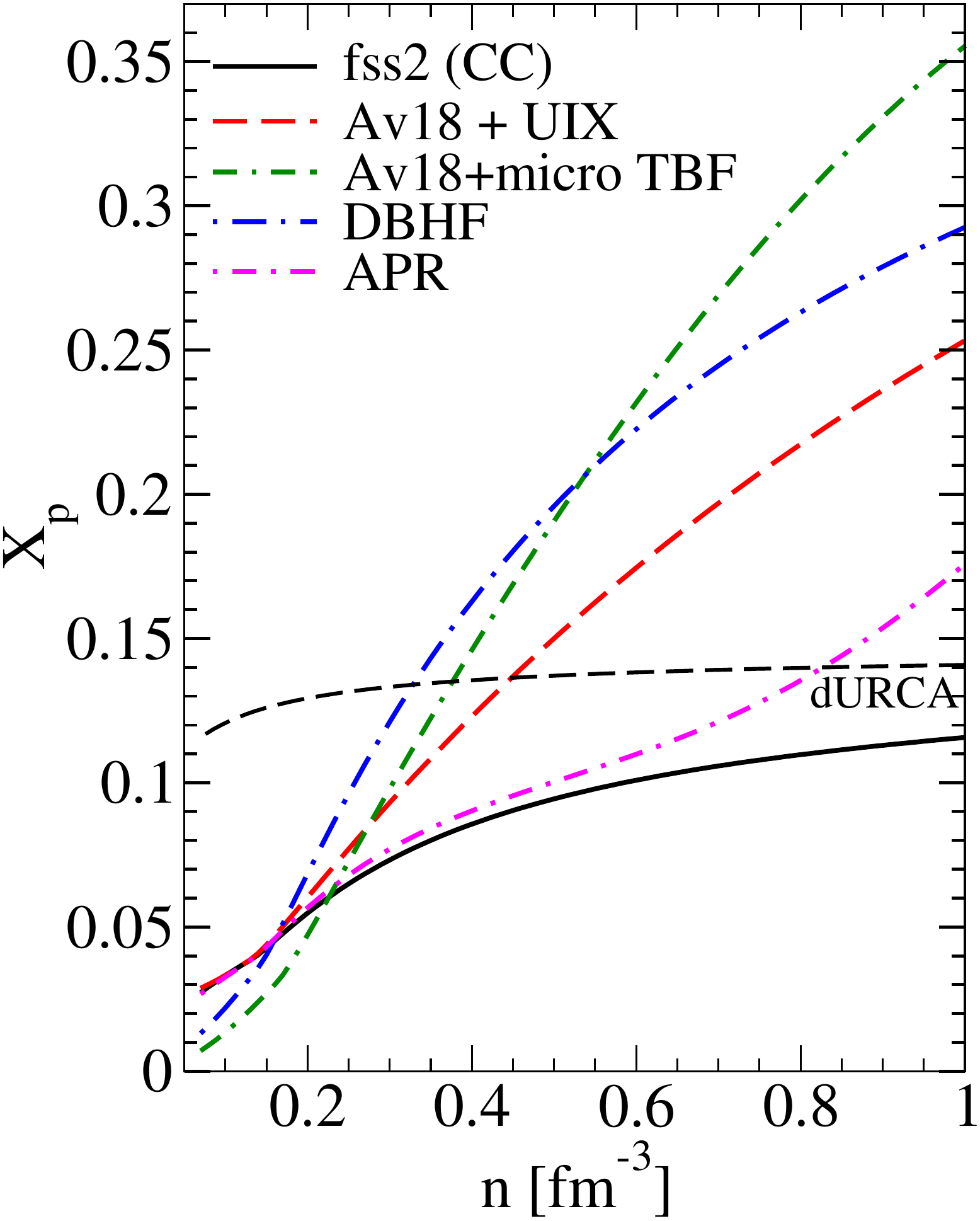}
\includegraphics[scale=0.35]{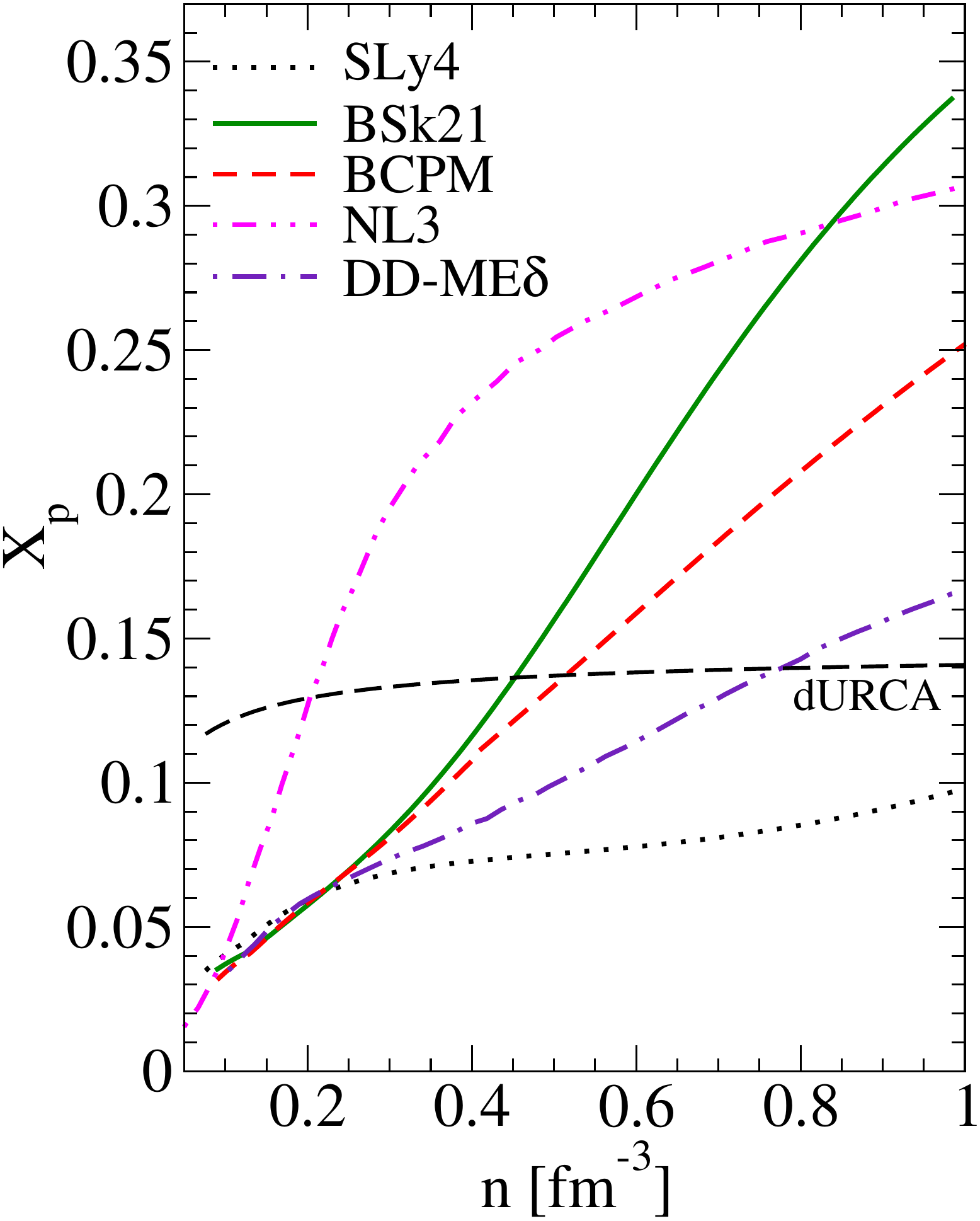}
\caption{Proton fraction versus baryon number density for different microscopic (left panel) and phenomenological (right panel) models. The dashed black lines labelled ``dURCA'' mark the threshold for the DU process to occur \cite{klahn2006}.}
\label{fig:xp}
\end{figure}

During the first $10^5-10^6$~yr, a NS cools down mainly via neutrino emission.
In the absence of superfluidity, three main processes are usually taken into account: the direct URCA (DU), the modified URCA (MU), and the NN bremsstrahlung (BNN) processes. 
The most efficient neutrino emission is the DU process, a sequence of neutron decays, $n \rightarrow p + e^- + \overline{\nu}_e$, and electron captures, $p+e^- \rightarrow n+\nu_e$.
For this process and for $npe$ NS matter, the neutrino emissivity is given by \cite{n_emis}
\begin{equation}
 Q^{(DU)} \approx
 4.0\times10^{27} \left(\frac{Y_e\ n_B}{n_0}\right)^{1/3} \frac{m_n^\star m_p^\star}{m_n^2} T_9^6 \mathop\Theta(k_{F_p} + k_{F_e} - k_{F_n})\ \mathrm{erg\ cm^{-2}\ s^{-1}} \:,
\label{e:emis}
\end{equation}
where $m_n$ is the neutron mass, $m_n^\star$ ($m_p^\star$) is the neutron (proton) effective mass, $T_9$ is the temperature in units of $10^9$~K, $\mathop\Theta$ is the Fermi function, and $k_{F,p}$, $k_{F,e}$, and $k_{F,n}$ are the proton, electron, and neutron Fermi momenta, respectively.
It thus considerably depends on the temperature and on the nucleon effective masses.
If muons are present, then the corresponding DU process may also become possible, in which case the neutrino emissivity is increased by a factor of 2.
If it takes place, the DU process enhances neutrino emission and NS cooling rates by a large factor compared to MU and BNN processes. 
The role of the DU processes has been long questioned in the past years, since it depends on the adopted EoS and the values of the superfluidity gaps, on which, at present, there is no consensus. 
Concerning the EoS, the energy and momentum conservation imposes a proton fraction threshold for this process to occur \cite{1991PhRvL..66.2701L, klahn2006}, $X_p \approx 11-15\%$, that is mainly determined by the symmetry energy.
In Fig.~\ref{fig:xp}, we display the proton fraction versus the baryon density in NS matter for different microscopic (left panel) and phenomenological (right panel) models.
For the former models, we observe that, except fss2 (CC), all the considered microscopic approaches are characterised by a quite low value of the threshold density.
For instance, for the BHF with A$v18 +$~UIX, the DU process sets in at $0.44$~fm$^{-3}$, thus the DU process operates in NS with masses $M > 1.10\,\ M_{\odot}$, while for the APR EoS the onset of DU is shifted to larger density, $0.82$~fm$^{-3}$, due to the lower values of the symmetry energy, hence $X_p$. 
Among the phenomenological models considered here, only the EoS based on the SLy4 EDF forbids the DU process, while the EoS based on NL3 (DD-ME$\delta$) has the lowest \cite{cav2011} (highest \cite{wang2014}) threshold density.
For the EoS based on the BSk21 (BCPM) EDF, the threshold density is $0.45$~fm$^{-3}$ \cite{fcpg2013} ($0.53$~fm$^{-3}$ \cite{sharma2015}), thus the DU process occurs for NS with $M > 1.59 \msun$ ($M>1.35 \msun$).
Incidentally, Kl\"{a}hn et al.~\cite{klahn2006} argued that no DU process should occur in NSs with typical masses in the range $M \sim 1-1.5 \msun$.
From the observational point of view, the pulsar in CTA1, the transiently accreting millisecond pulsar SAX J1808.4$-$3658, and the soft X-ray transient 1H 1905$+$000 appear to be very cold, thus suggesting that these NSs may cool very fast via the DU process \cite{jonker2007, heinke2009, page2009, abdo2012}. 
Moreover, the low luminosity from several young SN remnants likely to contain a still unobserved NS \cite{kaplan2004, kaplan2006} could suggest further evidence for a DU process \cite{shtyak2008, page2009}. 
If DU processes actually occur in those objects and the NS masses were known, they could put constraints on the EoSs unfavouring those that forbid DU for those masses. 
In fact, a key parameter that could discriminate whether the DU occurs in a NS is its mass.
Unfortunately, the masses of these cooling objects are not precisely measured, if not known at all.
An object of particular interest is Cassiopeia A \cite{heinkeho2010}, that can potentially give information on the interior of the NS (see, e.g., \cite{page2011, sht2011, bgv2012, bgv2013, sed2013, tar2016}) and on the nuclear symmetry energy and the nuclear pasta \cite{new2013}. 
Its fast cooling was claimed to be a direct proof of superfluidity in NSs, even if more recent analyses put a word of caution on the initial data \cite{poss2013, els2013, ho2015}.

A further critical point of most current cooling simulations is the fact that a given EoS is combined with pairing gaps obtained within a different theoretical framework and using different input interactions, thus resulting in an inconsistent analysis. 
Recently, some progress has been made along this direction \cite{tar2016}, concluding that the possibility of strong DU processes cannot be excluded from the cooling analysis. 

The current results confirm the extreme difficulty to draw quantitative conclusions from the current NS cooling data. 
In particular, the present substantial theoretical uncertainty regarding superfluidity gaps and thermal conductivity (see also Chap.~8 in this book) calls for a renewed effort in the theoretical activity of the next few years.

\section{Conclusions}
\label{sec:conclusions}


The EoS of hot and dense matter is a crucial input to describe static and dynamical properties of compact objects.
However, constructing such a (unified) EoS is a very challenging task.
The physical conditions prevailing in these astrophysical objects are so extreme that it is currently impossible to reproduce them in terrestrial laboratories.
Therefore, theoretical models are required.
Nevertheless, ab-initio calculations cannot be at present applied to determine the EoS in all the regions of NSs and SNe, mainly because of computational cost, thus more phenomenological models have to be employed.
In this Chapter, we have reviewed the current status of the EoS for compact objects.
We have presented the different underlying many-body methods, both microscopic and phenomenological, for homogeneous and inhomogeneous matter, considering only nucleonic degrees of freedom.
We have discussed these models with respect to constraints coming from both nuclear physics experiments and astrophysical observations: apart from the precise measurements of the $2 \msun$ NSs, other constraints are less strict since often model dependent.
New terrestrial experiments and facilities such as RIKEN, FAIR, HIE-ISOLDE, SPIRAL 2, FRIB, and TRIUMF, and new-generation telescopes and projects such as ATHENA+, NICER, and SKA, and gravitational-wave detectors such as Advanced Virgo and LIGO, and LISA promise to provide more and more precise data that can significantly contribute to probe the internal structure of compact objects, allowing unprecedented comparisons with theoretical predictions.
Finally, we have discussed some of the present challenges in the EoS modelling.
Indeed, despite many recent advances in the many-body treatment, still issues have to be faced in the description of the EoS.
These include (i) the model dependence of the constraints inferred from experimental nuclear physics and astrophysical data, (ii) the lack of a consistent and rigorous many-body treatment both at zero and finite temperature of cluster formation beyond the single-nucleus approximation, (iii) the treatment and role of nucleonic TBFs, and (iv) the description of the cooling in NSs.
The current theoretical uncertainties require significant efforts to be undertaken in these directions in the next few years.

\begin{acknowledgement}
This work has been partially supported by the COST action MP1304 ``NewCompStar''.
The authors would like to thank Ad. R. Raduta for providing Fig.~4, H.~J. Schulze for providing Fig.~1, M. Baldo for providing us with data for Fig. 2 and for valuable comments on the manuscript, N. Chamel and S. Goriely for fruitful discussions and providing us with data for Figs.~3, 5, 6, C. Provid{\^e}ncia for data for Figs.~3, 5, F. Gulminelli for valuable comments on the manuscript, and X. Roca Maza for insightful discussions.
\end{acknowledgement}

\bibliographystyle{spmpsci}
\bibliography{biblio_ch6}

\end{document}